%% file: YukITAMSNoisyChannel2012FinalSubmissionv5.tex
\documentclass[journal]{IEEEtran}

\usepackage{amsmath,amssymb,latexsym, bm,bbm,mathrsfs}
\usepackage{epsfig}
\usepackage{caption}

\input{symbols}

\def\Ebox#1#2{%
\medskip
\begin{center}
  \strut\epsfxsize=#1 \hsize\epsfbox{#2}
\end{center}
\smallbreak}

\def\stp{{\cal T}}
\def\sampd{\mathop{\hbox{\rm J}}}

\newlength{\noteWidth}
\long\def\notes#1{\ifinner
           {\tiny #1}
           \else
           \marginpar{\parbox[t]{\noteWidth}{\raggedright\tiny #1}}
       \fi\typeout{#1}}

       \newtheorem{thm}{\bf{Theorem}}[section]
       \newtheorem{cor}{\bf{Corollary}}[section]
       \newtheorem{lem}{\bf{Lemma}}[section]

       \newtheorem{defn}{\bf{Definition}}[section]
  \newtheorem{remark}{\bf{Remark}}[section]

\newcounter{mytempeqncnt}

\fussy

\begin{document}

\title{Characterization of Information Channels for Asymptotic Mean Stationarity and Stochastic Stability of Non-stationary/Unstable Linear Systems}

\author{Serdar Y\"uksel$^1$}

\maketitle
\footnotetext[1]{Department of Mathematics and Statistics, Queen's University, Kingston, Ontario, Canada, K7L 3N6. Research supported by the Natural Sciences and Engineering Research Council of Canada (NSERC). Email: yuksel@mast.queensu.ca. This work was presented in part at the Annual Workshop on Information Theory and Applications, at the University of California San Diego, and at the IEEE Conference on Decision and Control, Florida, in 2011.\\
Copyright (c) 2011 IEEE. Personal use of this material is permitted.  However, permission to use this material for any other purposes must be obtained from the IEEE by sending a request to pubs-permissions@ieee.org.}
\begin{abstract}
Stabilization of non-stationary linear systems over noisy communication channels is considered. Stochastically stable sources, and unstable but noise-free or bounded-noise systems have been extensively studied in information theory and control theory literature since 1970s, with a renewed interest in the past decade. There have also been studies on non-causal and causal coding of unstable/non-stationary linear Gaussian sources. In this paper, tight necessary and sufficient conditions for stochastic stabilizability of unstable (non-stationary) possibly multi-dimensional linear systems driven by Gaussian noise over discrete channels (possibly with memory and feedback) are presented. Stochastic stability notions include recurrence, asymptotic mean stationarity and sample path ergodicity, and the existence of finite second moments. Our constructive proof uses random-time state-dependent stochastic drift criteria for stabilization of Markov chains. For asymptotic mean stationarity (and thus sample path ergodicity), it is sufficient that the capacity of a channel is (strictly) greater than the sum of the logarithms of the unstable pole magnitudes for memoryless channels and a class of channels with memory. This condition is also necessary under a mild technical condition. Sufficient conditions for the existence of finite average second moments for such systems driven by unbounded noise are provided.
\end{abstract}


\section{Problem Formulation}

This paper considers stochastic stabilization of linear systems controlled or estimated over discrete noisy channels with feedback. We consider first a scalar LTI discrete-time system (we consider multi-dimensional systems in Section \ref{MultiDimensionalSection}) described by
\begin{eqnarray}
\label{ProblemModel4}
x_{t+1}=ax_{t} + bu_{t} + d_t, \quad \quad t \geq 0
\end{eqnarray}
Here $x_t$ is the state at time $t$, $u_t$ is the control input, the initial condition $x_0$ is a second order random variable, and $\{d_t \}$ is a sequence of zero-mean independent, identically distributed (i.i.d.) Gaussian random variables. 
It is assumed that     $|a| \geq 1$ and $b \neq 0$:  The system is open-loop unstable, but it is stabilizable.

\begin{figure}[h]
\centering
\Ebox{.95}{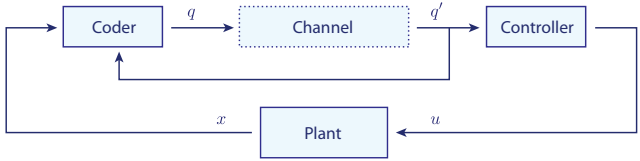}
\caption{Control over a discrete noisy channel with feedback. \label{LLL}}
\end{figure}

This system is connected over a Discrete Noisy Channel with a finite capacity to a controller, as shown in Figure~\ref{LLL}.

The controller has access to the information it has received through the channel. The controller in our model estimates the state and then applies its control.

\begin{remark}
We note that the existence of the control can also be regarded as an estimation correction, and all results regarding stability may equivalently be viewed as the stability of the estimation error. Thus, the two problems are identical for such a controllable system and the reader unfamiliar with control theory can simply replace the stability of the state, with the stability of the estimation error.
\end{remark}

Recall the following definitions.
\begin{defn}
A finite-alphabet channel with memory is characterized by a sequence of finite input
alphabets ${\cal M}^{n+1}$, finite output alphabets ${\cal M'}^{n+1}$, and a sequence of conditional probability measures $P_n(q'_{[0,n]}|q_{[0,n]})$, from ${\cal M}^{n+1} \times {\cal M'}^{n+1}$ to $\mathbb{R}$, with,
\[q'_{[0,n]}=:\{q'_0,q'_1,\dots,q'_n\} \quad \quad q_{[0,n]}:=\{q_0,q_1,\dots,q_n\}.\]
\end{defn}

\begin{defn}
A Discrete Memoryless Channel (DMC) is characterized by a finite input
alphabet ${\cal M}$, a finite output alphabet ${\cal M}'$, and
a conditional probability mass function $P(q'|q)$, from ${\cal M} \times {\cal M'}$ to $\mathbb{R}$. Let
 $q_{[0,n]} \in {\cal M}^{n+1}$ be a sequence of input symbols, and let $q'_{[0,n]} \in {\cal M'}^{n+1}$ be a sequence of output symbols, where $q_k \in {\cal M}$ and $q'_k \in {\cal M'}$ for all $k$. Let
 $P^{n+1}_{DMC}$ denote the joint mass function on the $n+1$-tuple  input  and output spaces. A DMC from ${\cal M}^{n+1}$ to ${\cal M'}^{n+1}$ satisfies the following: $P^{n+1}_{DMC}(q'_{[0,n]},q_{[0,n]})=\prod_{k=0}^{n} P_{DMC}(q'_k,q_k),$ $\forall q_{[0,n]}\in {\cal M}^{n+1}$,
$q'_{[0,n]} \in {\cal M'}^{n+1}$, where $q_k,q'_k$ denote the $k$th component of the vectors $q_{[0,n]}, q'_{[0,n]}$, respectively. \qed
\end{defn}

In the problem considered, a source coder maps the information at the encoder to corresponding channel inputs. This is done through quantization and a channel encoder.
The quantizer outputs are transmitted through a channel, after being subjected to a channel encoder. The receiver has access to noisy versions of the quantizer/coder outputs for each time, which we denote by $q'_t \in {\cal M}'$.
The quantizer and the source coder policy is causal such that the channel input at time $t \geq 0$, $q_t$, is generated using the information vector $I^s_t$ available at the encoder for $t > 0$:
$$I^s_t = \{I^s_{t-1},x_t, q_{t-1}, q'_{t-1} \},$$
and $I^s_0=\{\nu_0,x_0\}$, where $\nu_0$ is the probability measure for the initial state. 

The control policy at time $t$, also causal, is measurable on the sigma-algebra generated by $I^c_t$, for $t\geq 1$:
$$I^c_t = \{I^c_{t-1}, q'_t \},$$ and $I^c_0=\{\nu_0\},$
and is a mapping to $\mathbb{R}$.

We will call such coding and control policies {\em admissible} policies.

The goal of the paper is to identify conditions on the channel under which the controlled process $\{x_t\}$ is stochastically stable in sense that $\{x_t\}$ is recurrent, $\{x_t\}$ is asymptotically mean stationary and satisfies Birkhoff's sample path ergodic theorem, and that $\lim_{T \to \infty} {1 \over T} \sum_{t=0}^{T-1}||x_t||^2$ is finite almost surely, under admissible coding and control policies. We will make these notions and the contributions of the paper more precise after we discuss a literature review in the next section. The appendix in Section \ref{sectionergodicandMarkovreview} contains a review of relevant definitions and results on stochastic stability of Markov chains and ergodic theory.

Here is a brief summary of the paper. In the following, we will first provide a comprehensive literature review. In Section II, we state the main results of the paper. In Section III, we consider extensions to channels with memory, and in Section IV, we consider multi-dimensional settings. Section V contains the proofs of necessity and sufficiency results for stochastic stabilization. Section VI contains concluding remarks and discusses a number of extensions. The paper ends with an appendix, in Section \ref{sectionergodicandMarkovreview}, which contains a review of stochastic stability of Markov chains and a discussion on ergodic processes.

\subsection{Literature Review}

There is a large literature on stochastic stabilization of sources via coding, both in the information theory and control theory communities.

In the information theory literature, stochastic stability results are established mostly for stationary sources, which are already in some appropriate sense stable sources. In this literature, the stability of the estimation errors as well as the encoder state processes are studied. These systems mainly involve causal and non-causal coding (block coding, as well as sliding-block coding) of stationary sources \cite{KiefferDunham}, \cite{GoodmanGersho}, \cite{Kieffer}, and asymptotically mean stationary sources \cite{Gray2}. Real-time settings such as sigma-delta quantization schemes have also been considered in the literature, see for example \cite{Gray3} among others.

There also have been important contributions on non-causal coding of non-stationary/unstable sources: Consider the following Gaussian AR process:
\begin{eqnarray}\label{ARMA}
x_t = - \sum_{k=1}^{m} a_k x_{t-k} + w_k,
\end{eqnarray}
where $\{w_k\}$ is an independent and identical, zero-mean, Gaussian random sequence with variance $E[w_1^2]= \sigma^2$. If the roots of the polynomial: ${\cal P}(z):=1 + \sum_{k=1}^m a_k z^{-k}$ are all in the interior of the unit circle, then the process is stationary and its rate distortion function (with the distortion being the expected, normalized Euclidean error) is given parametrically (in terms of parameter $\theta$) by the following Kolmogorov's formula \cite{Kolmogorov} \cite{Gray70}, obtained by considering the asymptotic distribution of the eigenvalues of the correlation matrix:
\[D_{\theta} = {1 \over 2 \pi} \int_{-\pi}^{\pi} \min(\theta, {1 \over g(w)}) dw,\]
\[R(D_{\theta})  = {1 \over 2 \pi} \int_{-\pi}^{\pi} \max({1 \over 2} (\log{1 \over \theta g(w)}), 0) dw,\]
with $g(w) = {1 \over \sigma^2} |1+\sum_{k=1} a_k e^{-ik w}|^2$.
If at least one root, however, is on or outside the unit circle, the analysis is more involved as the asymptotic eigenvalue distribution contains unbounded components. \cite{Hashimoto}, \cite{Gray70} and \cite{GrayHashimoto} showed that, using the properties of the eigenvalues as well as Jensen's formula for integrations along the unit circle, $R(D_{\theta})$ above should be replaced with:
\begin{eqnarray}\label{hashimoto}
R(D_{\theta}) &&= {1 \over 2 \pi} \int_{-\pi}^{\pi} \max\bigg({1 \over 2} \log({1 \over \theta g(w)}), 0\bigg) dw \nonumber \\
&& \quad \quad \quad \quad + \sum_{k=1}^m {1 \over 2} \max\bigg(0,\log(|\rho_k|^2)\bigg),
\end{eqnarray}
where $\{\rho_k\}$ are the roots of the polynomial ${\cal P}$. We refer the reader to a review in \cite{GrayHashimoto} regarding rate-distortion results for such non-stationary processes and on the methods used in \cite{Gray70} and \cite{Hashimoto}.

Reference \cite{BergerWiener} obtained the rate-distortion function for Wiener processes, and in addition, developed a two-part coding scheme, which was later generalized for more general processes in \cite{SahaiISIT04} and \cite{SahaiParts}, which we will discuss below further, to unstable Markov processes. The scheme in \cite{BergerWiener} exploits the independent increment property of Wiener processes.

Thus, an important finding in the above literature is that, the logarithms of the unstable poles in such linear systems appear in the rate-distortion formulations, an issue which has also been observed in the networked control literature, which we will discuss further below. We also wish to emphasize that these coding schemes are non-causal, that is the encoder has access to the entire ensemble before the encoding begins. 

In contrast with information theory, due to the practical motivation of sensitivity to delay, the control theory literature has mainly considered causal/zero-delay coding for unstable (or non-stationary) sources, in the context of networked control systems. In the following, we will provide a discussion on the contributions in the literature which are contextually close to our paper.

 \cite{BorkarMitter} studied  the trade-off between delay and reliability, and posed questions leading to an accelerated pace of research efforts on what we today know as networked control problems. References \cite{Brockett}, \cite{Tatikonda}, and \cite{NairEvans} obtained the minimum lower bound needed for stabilization over noisy channels under a class of assumptions on the system noise and channels. This result states that for stabilizability under information constraints, in the mean-square sense, a minimum rate needed for stabilizability has to be at least the sum of the logarithms of the unstable poles/eigenvalues in the system; that is:
\begin{eqnarray} \label{dataRateTheorem}
\sum_{k=1}^m {1 \over 2} \max\bigg(0,\log(|\rho_k|^2)\bigg).
\end{eqnarray}
Comparing this result with (\ref{hashimoto}), we observe that, the rate requirement is not due to causality but to the (differential) entropy rate of the unstable system.

For coding and information transmission for unstable linear systems, there is an important difference between continuous alphabet and finite-alphabet (discrete) channels as discussed in \cite{YukselBasarTAC2011}: When the space is continuous alphabet, we do not necessarily need to consider adaptation in the encoders. On the other hand, when the channel is finite alphabet, and the system is driven by unbounded noise, a static quantizer leads to almost sure instability (see Proposition 5.1 in \cite{NairEvans} and Theorem 4.2 in \cite{YukselBasarTAC2011}). With this observation, \cite{NairEvans} considered a class of variable rate quantizer policies for such unstable linear systems driven by noise, with unbounded support set for its probability measure, controlled over noiseless channels, and obtained necessary and sufficient conditions for the boundedness of the following expression
\[\limsup_{t \to \infty} E[||x_t||^2] < \infty.\]
With fixed rate, reference \cite{YukTAC2010} obtained a somewhat stronger expression and established a limit
\[\lim_{t \to \infty} E[||x_t||^2] < \infty,\]
and obtained a scheme which made the state process and the encoder process stochastically stable in the sense that the joint process is a positive Harris recurrent Markov chain and the sample path ergodic theorem is applicable.

Reference \cite{Martins2} established that when a channel is present in a controlled linear system, under stationarity assumptions, the rate requirement in (\ref{dataRateTheorem}) is necessary for having finite second moments for the state variable. A related argument was made in \cite{YukselBasarTAC2011} under the assumption of invariance conditions for the controlled state process under memoryless policies and finite second moments. In this paper, in Theorem \ref{StochasticStabilityofVectorNecessity}, we will present a very general result along this direction for a general class of channels and a weaker stability notion. Such settings were further considered in the literature. The problem of control over noisy channels has been considered in many publications including \cite{BansalBasar}, \cite{Tatikonda}, \cite{SahaiParts}, \cite{Martins}, \cite{SavkinSIAM09}, \cite{MatveevSavkin}, \cite{Minero}, \cite{TatikondaSahaiMitter} among others. Many of the constructive results involve Gaussian channels, or erasure channels (some modeled as infinite capacity erasure channels as in \cite{Schenato} and \cite{ImerYukselBasar}). Other works have considered cases where there is either no disturbance or that the disturbance is bounded, with regard to noisy sources and noisy channels. We discuss some of these in the following.

It is to be stressed that, the notion of stochastic stability is very important in characterizing the conditions on the channel. References \cite{SavkinSIAM09}, \cite{MatveevSavkin} considered stabilization in the following sense, when the system noise is bounded:
\[\limsup_{t \to \infty} |x_t| <  \infty \quad a.s.,\]
and observed that one needs the zero-error capacity (with feedback) to be greater than a particular lower bound. A similar observation was made in \cite{SahaiParts}, which we will discuss further in the following. When the system is driven by noise which admits a probability measure with unbounded support, the stability requirement above is impossible for an infinite horizon problem, even when the system is open-loop stable, since for any bound, there exists almost surely a realization of a noise variable which will be larger.

References \cite{Sahai}, \cite{SahaiParts} considered systems driven by bounded noise and considered a number of stability criteria: Almost sure stability for noise-free systems, moment stability for systems with bounded noise ($\limsup_{t \to \infty} E[|x_t|^p] <  \infty$) as well as {\em stability in probability} (defined in \cite{MatveevSavkin}) for systems with bounded noise. Stability in probability is defined as follows: For every $p>0$, there exists a $\zeta$ such that $P(|x_t| > \zeta) < p$ for all $t \in \mathbb{N}$. \cite{Sahai} and \cite{SahaiParts} also offered a novel and insightful characterization for reliability for controlling unstable processes, named, any-time capacity, as the characterization of channels for which the following criterion can be satisfied:
\[\limsup_{t \to \infty} E[|x_t|^p] < \infty,\]
for positive moments $p$. A channel is $\alpha-$ any-time reliable for a sequential coding scheme if: $P(\hat{m}^{t-d}(t) \neq m^{t-d}(t)) \leq K 2^{-\alpha d}$ for all $t,d$. Here $m^{t-d}$ is the message transmitted at time $t-d$, estimated at time $t$. One interesting aspect of an any-time decoder is the independence from the delay, with a fixed encoder policy. \cite{SahaiParts} states that for a system driven by bounded noise, stabilization is possible if the maximum rate for which an any-time reliability of $2\log_2(|\rho_1|)$ is satisfied, is greater than $\log_2(|\rho_1|)$, where $\rho_1$ is the unstable pole of a linear system.

In a related context, \cite{Martins}, \cite{SahaiParts}, \cite{MatveevSavkin} and \cite{Matveev} considered the relevance to Shannon capacity. \cite{Martins} observed that when the moment coefficient goes to zero, Shannon capacity provides the right characterization on whether a channel is sufficient or insufficient, when noise is bounded. A parallel argument is provided by \cite{SahaiParts}, in Section III.C.1, observing that in the limit when $p \to 0$, capacity should be the right measure for the objective of satisfying {\em stability in probability}. Their discussion was for bounded noise signals. \cite{MatveevSavkin} also observed a parallel discussion, again for bounded noise signals.

With a departure from the bounded noise assumption, \cite{Matveev} extended the discussion in \cite{SahaiParts} and studied a more general model of multi-dimensional systems driven by an unbounded noise process considering again {\em stability in probability}. \cite{Matveev} also showed that when the discrete noisy channel has capacity less than $\log_2(|a|)$, where $a$ is defined in (\ref{ProblemModel4}), there exists no stabilizing scheme, and if the capacity is strictly greater than this number, there exists a stabilizing scheme in the sense of stability in probability.

Many network applications and networked control applications require the access of control and sensor information to be observed intermittently. Toward generating a solution for such problems, \cite{YukMeynTAC2010} and \cite{YukselAllerton09} developed random-time state-dependent drift conditions leading to the existence of an invariant distribution possibly with moment constraints, extending the earlier deterministic state dependent results in \cite{MeynStateDrift}. Using drift arguments, \cite{YukselBasarTAC2011} considered noisy (both discrete and continuous alphabet) channels, \cite{YukTAC2010} considered noiseless channels and \cite{YukMeynTAC2010} considered erasure channels for the following stability criteria: The existence of an invariant distribution, and the existence of an invariant distribution with finite moments.

References \cite{Elia}, \cite{Martins2} and \cite{Martins3} considered general channels (possibly with memory), and with a connection with Jensen's formula and Bode's sensitivity integral, developed necessary and sufficient rates for stabilization under various networked control settings. Reference \cite{Minero} considered erasure channels and obtained necessary and sufficient time-varying rate conditions for control over such channels. Reference \cite{Coviello} considered second moment stability over a class of Markov channels with feedback and developed necessary and sufficient conditions, for systems driven by an unbounded noise. Reference \cite{Gurt} considered the stochastic stability over erasure channels, parallel to the results of \cite{YukMeynTAC2010}.

On the other hand, for more traditional information theoretic settings where the source is revealed at the beginning of transmission, and for cases where causality and delay are not important, the separation principle for source and channel coding results are applicable for ergodic sources and information stable channels. The separation principle for more general setups has been considered in \cite{VerduSeparation}, among others. References \cite{WitsenhausenReal} and \cite{WalrandVaraiya} studied the optimal causal coding problem over respectively a noiseless channel  and  a noisy channel with noiseless feedback. Unknown sources have been considered in \cite{Charalambous}. We also note that, when noise is bounded, binning based strategies, inspired from Wyner-Ziv and Slepian-Wolf coding schemes are applicable. This type of consideration has been applied in \cite{SahaiParts}, \cite{YukselBasarTAC2011} and \cite{Pulkit}. Finally quantizer design for noiseless or bounded noise systems include \cite{Zampieri}, \cite{EliaMitter} and \cite{IshiiFrancis}. Channel coding algorithms for control systems have been presented recently in \cite{SimsekTAC}, \cite{Schulman} and \cite{Ravi}.

There also has been progress on coding for noisy channels for the transmission of sources with memory. Due to practical relevance, for communication of sources with memory over channels, particular emphasis has been placed on Gaussian channels with feedback. For such channels, the fact that real-time linear schemes are rate-distortion achieving has been observed in \cite{Berger}, \cite{Gastpar}; and \cite{BansalBasar} in a control theoretic context. Aside from such results (which involve matching between rate-distortion achieving test channels and capacity achieving source distributions \cite{Gastpar}), capacity is known not to be a good measure of information reliability for channels for real-time (zero-delay or delay sensitive) control and estimation problems, see \cite{WalrandVaraiya} and \cite{SahaiParts}. Such general aspects of {\em comparison of channels} for cost minimization have been investigated in \cite{Blackwell} among others.

Also in the information theory literature, performance of information transmission schemes for channels with feedback has been a recurring avenue of research in information theory, for both variable length and fixed length coding schemes \cite{Hornstein}, \cite{Burnashev}, \cite{SahaiDelayBlock}, \cite{DraperSahai}. In such setups, the source comes from a fixed alphabet, except the sequential setup in \cite{SahaiDelayBlock} and \cite{DraperSahai}.

\subsection{Contributions of the Paper}

In view of the discussion above, the paper makes the following contributions. The question: {\em When does a linear system driven by unbounded noise, controlled over a channel (possibly with memory) satisfy Birkhoff's sample path ergodic theorem (or is asymptotically mean stationary)?}, has not been answered to our knowledge. Also, the finite moment conditions for an arbitrary discrete memoryless channel for a system driven by unbounded noise have not been investigated to our knowledge, except for the bounded noise analysis in \cite{SahaiParts}. The contributions of the paper are on the two problems stated above. In this paper, we will show that, the results in the literature can be strengthened to asymptotic mean stationarity and ergodicity. As a consequence of Kac's Lemma \cite{Cover}, {\em stability in probability} can also be established. We will also consider conditions for finite second moment stability. We will use the random-time state-dependent drift approach \cite{YukMeynTAC2010} to prove our achievability results. Hence, we will find classes of channels under which the controlled process $\{x_t\}$ is stochastically stable in in each of the following senses:
\begin{itemize}
\item $\{x_t\}$ is recurrent: There exists a compact set $A$ such that $\{x_t \in A \}$ infinitely often almost surely.
\item $\{x_t\}$ is asymptotically mean stationary and satisfies Birkhoff's sample path ergodic theorem. We will establish that Shannon capacity provides a boundary point in the space of channels on whether this objective can be realized or not, provided a mild technical condition holds.
\item $\lim_{T \to \infty} {1 \over T} \sum_{t=0}^{T-1}||x_t||^2$ exists and is finite almost surely.
\end{itemize}

\section{Stochastic Stabilization over a DMC}

\subsection{Asymptotic Mean Stationarity and $n-$ergodicity}

\begin{thm}\label{NecessityRecurrence}
For a controlled linear source given in (\ref{ProblemModel4}) over a DMC under any admissible coding and controller policy, to satisfy the AMS property under the following condition
\[\liminf_{T \to \infty} {1 \over T} h(x_T) \leq 0, \]
the channel capacity $C$ must satisfy
\[C \geq \log_2(|a|).\]
\qed
\end{thm}

\textbf{Proof:} See the proof of Theorem \ref{StochasticStabilityofVectorNecessity} in Section \ref{proofNecessity}

\begin{remark}
The condition $\liminf_{T \to \infty} {1 \over T} h(x_T) \leq 0$ is a very weak condition. For example a stochastic process whose second moment grows subexponentially in time such that \[\liminf_{T \to \infty} {\log(E[x_T^2]) \over T} = 0\] satisfies this condition.
\end{remark}

The above condition is almost sufficient as well, as we state in the following.

\begin{thm}\label{RecurrenceChannelNoAssumption}
For the existence of a compact coordinate recurrent set (see Definition \ref{CoordinateRecurrent}), the following is sufficient: The channel capacity $C$ satisfies: $C > \log_2(|a|)$.
\qed \end{thm}

\textbf{Proof:} See Section \ref{SufficiencyRecurrence}. \qed

For the proof, we consider the following update algorithm. The algorithm and its variations have been used in source coding and networked control literature: See for example the earlier papers \cite{GoodmanGersho}, \cite{Kieffer}, and more recent ones \cite{BrockettLiberzon}, \cite{NairEvans}, \cite{SavkinSIAM09},  \cite{MatveevSavkin}, \cite{Matveev}, \cite{YukTAC2010} and \cite{YukTutorial2011}. Our contribution is primarily on the stability analysis.

\begin{figure}[h]
\centering
\epsfig{figure=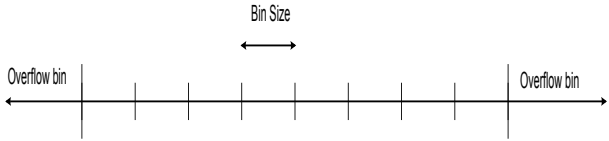,height=1.8cm,width=8cm}
\caption{A uniform quantizer with a single overflow bin. \label{LLLL}}
\end{figure}

Let $n$ be a given block length. Consider the following setup. We will consider a class of uniform quantizers, defined by two parameters, with bin size
$\Delta>0$, and an even number $K(n) \ge 2$ (see Figure \ref{LLLL}).
The uniform quantizer map  is defined as follows:  For $k=1,2\dots,K(n)$,

\begin{eqnarray}\label{UniformQuantizerRule}
Q_{K(n)}^{\Delta}(x) = \begin{cases}   (k - \half (K(n) + 1) ) \Delta,
	 \nonumber \\
  \mbox{if} \quad x \in [ (k-1-\half K(n)  ) \Delta , (k-\half K(n)  ) \Delta) \nonumber \\
  (\half (K(n) - 1) ) \Delta,  \quad \mbox{if} \quad  x = \half K(n) \Delta \nonumber \\
   {\cal Z} , \quad \quad \quad \mbox{if} \quad |x| > \half K(n) \Delta.
\end{cases} \nonumber
\end{eqnarray}
where ${\cal Z}$ denotes the overflow symbol in the quantizer. We define $\{x: Q_{K(n)}^{\Delta}(x) \neq {\cal Z}\}$ to be the {\em granular region} of the quantizer.


At every sampling instant $t=kn, k=0,1,2,\dots$, the source coder ${\cal E}^s_t$ quantizes output symbols in $\mathbb{R} \cup \{{\cal Z}\}$ to a set ${\cal M}(n) = \{1,2,\dots,K(n)+1\}$. A channel encoder ${\cal E}^c_t$ maps the elements in ${\cal M}(n)$ to corresponding channel inputs $q_{[kn,(k+1)n-1]} \in {\cal M}^n$.



For each time $t =kn-1, k =1,2,3,\dots$, the channel decoder applies a mapping ${\cal D}_{tn}: {\cal M'}^{n} \to {\cal M}(n)$, such that \[c'_{(k+1)n-1}={\cal D}_{kn}(q'_{[kn,(k+1)n-1]}).\]
Finally, the controller runs an estimator:
\begin{eqnarray*}
&&\hat{x}_{kn} = ({\cal E}^s_{kn})^{-1}(c'_{(k+1)n-1}) \times 1_{\{ c'_{(k+1)n-1} \neq {\cal Z}\}} \\
&& \quad \quad \quad \quad \quad \quad \quad \quad  + 0 \times 1_{\{c'_{(k+1)n-1}={\cal Z}\}},
\end{eqnarray*}

where $1_E$ denotes the indicator function for event $E$. Hence, when the decoder output is the overflow symbol, the estimation output is $0$.

We consider quantizers that are adaptive. In our setup, the bin size of the uniform quantizer acts as the state of the quantizer. At time ${kn}$ the bin size, $\Delta_{kn},$ is assumed to be a function of the previous state $\Delta_{(k-1)n}$ and the past $n$ channel outputs. We assume that the encoder has access to the previous channel outputs. Thus, such a quantizer is implementable at both the encoder and the decoder.

With $K(n) > \lceil |a|^{n} \rceil$, $R = \log_2(K(n) + 1)$, let us define $R'(n)=\log_2(K(n))$ and let
\[R'(n) > n \log_2({|a| \over \alpha}),\]
 for some $\alpha, 0 < \alpha < 1$ and $\delta >0$.
 When clear from the context, we will drop the index $n$ in $R'(n)$.

We will consider the following update rules in the controller actions and the quantizers.
For $t \geq 0$ and with $\Delta_0 > L$ for some $L \in \mathbb{R}_+$, and $\hat{x}_0 \in \mathbb{R}$, consider:
For $t=kn$, $k \in \mathbb{N}$
\begin{eqnarray}\label{QuantizerUpdate2}
 u_t &=& - 1_{\{t=(k+1)n-1\} }  {a^n \over b} \hat{x}_{kn}, \nonumber \\
 \Delta_{(k+1)n} &=& \Delta_{kn} \bar{Q}(\Delta_{kn}, c'_{(k+1)n-1}),
\end{eqnarray}
where $c'$ denotes the decoder output variable. If we use $\delta >0$ and $L > 0$ such that,
\begin{eqnarray}
 \bar{Q}(\Delta,c') = (|a| + \delta)^n \quad && \mbox{if } \quad c' = {\cal Z},   \nonumber \\
 \bar{Q}(\Delta,c') = \alpha^n \quad && \mbox{if } \quad c'  \neq {\cal Z}, \Delta \geq L,   \nonumber \\
 \bar{Q}(\Delta,c') = 1 \quad \quad  && \mbox{if } \quad c'  \neq {\cal Z}, \Delta < L,
\end{eqnarray}
we will show that a recurrent set exists. The above imply that $\Delta_t \geq L \alpha^n =:L'$ for all $t \geq 0$.

Thus, we have three main events: When the decoder output is the overflow symbol, the quantizer is zoomed out (with a coefficient of $(|a| + \delta)^n$). When the decoder output is not the overflow symbol ${\cal Z}$, the quantizer is zoomed in (with a coefficient of $\alpha^n$) if the current bin size is greater than or equal to $L$, and otherwise the bin size does not change.

We will establish our stability result through random-time stochastic drift criterion of Theorem \ref{thm5}, developed in \cite{YukMeynTAC2010} and \cite{YukselAllerton09}. This is because of the fact that, the quantizer helps reduce the uncertainty on the system state only when the state is in the {\em granular} region of the quantizer. The times when the state is in this region are random. The reader is referred to Section \ref{RandomDriftTheoremSection} in the appendix for a detailed discussion on the drift criteria.

In the following, we make the quantizer bin size process space countable and as a result establish the irreducibility of the sampled process $(x_{tn}, \Delta_{tn})$. 

\begin{thm}\label{Cyclostationary}
For an adaptive quantizer satisfying Theorem \ref{RecurrenceChannelNoAssumption}, suppose that the quantizer bin sizes are such that their logarithms are integer multiples of some scalar $s$,  and $\log_2(\bar{Q}( \varble ))$ takes values in integer multiples of $s$. Suppose the integers taken are relatively prime (that is they share no common divisors except for $1$). Then the sampled process $(x_{tn},\Delta_{tn})$ forms a positive Harris recurrent Markov chain at sampling times on the space of admissible quantizer bins and state values.  
\qed
\end{thm}

\textbf{Proof:} See Section \ref{ProofInv}. \qed

\begin{thm}\label{ErgodicityImpliedSampled}
Under the conditions of Theorems \ref{RecurrenceChannelNoAssumption} and \ref{Cyclostationary}, the process $\{x_t,\Delta_{t}\}$ is $n$-stationary, $n$-ergodic and hence AMS.
\qed \end{thm}

\textbf{Proof:} See Section \ref{ProofErgodicityImpliedSampled}. \qed

The proof follows from the observation that a positive Harris recurrent Markov chain is ergodic. It uses the property that if a sampled process is a positive Harris recurrent Markov chain, and if the intersampling time is fixed, with a time-homogenous update in the inter-sampling times, then the process is mixing, $n-$ergodic and $n-$stationary.

\subsection{Finite Second Moment}\label{sectionFiniteSecondMoment}

In this section, we discuss finite moment stability. Such an objective is important in applications. In control theory, quadratic cost functions are the most popular ones for linear and Gaussian systems. Furthermore, the two-part coding scheme of Berger in \cite{Berger} can be generalized for more general unstable systems if one can prove finite moment boundedness of sampled end-points. 

For a given coding scheme with block-length $n$ and a message set ${\cal M}(n)=\{1,2,\dots,K(n)+1\}$, and a decoding function $\gamma: {\cal M}'^{n} \to \{1,2,\dots,K(n)+1\}$ define three types of errors:

\begin{itemize}
\item Type I-A: Error from a granular symbol to another granular symbol. We define a bound for such errors. Define $P^e_{g|g}(n)$ to be
\[\max_{c \in {\cal M}(n) \setminus {\cal Z}} P(\gamma(q'_{[0,n-1]}) \neq c, \gamma(q'_{[0,n-1]}) \neq {\cal Z} | c),\]
where conditioning on $c$ means that the symbol $c$ is transmitted.
\item Type I-B: Error from a granular symbol to ${\cal Z}$: We define the following.
\[P^e_{g|g}(n): = \max_{c \in {\cal M}(n) \setminus {\cal Z}} P(\gamma(q'_{[0,n-1]}) = {\cal Z} | c)\]

\item Type II: Error from ${\cal Z}$ to a granular symbol:
\[P^e_{g|{\cal Z}}(n):= P(\gamma(q'_{[0,n-1]}) \neq {\cal Z} | {\cal Z}) \]
\end{itemize}

Type II error will be shown to be crucial in the analysis of the error exponent. Type I-A and I-B will be important for establishing the drift properties. We summarize our results below.

\begin{thm}\label{FiniteMomentChannel}
A sufficient condition for second moment stability (for the joint $(x_t,\Delta_t)$ process) over a discrete memoryless channel (DMC) is that:
\[\lim_{n \to \infty} ({1 \over n} \log(P^e_{{\cal Z}|g}(n)) + 2\log(|a|+\delta) < 0,\]
\[\lim_{n \to \infty} (\kappa {1 \over n} \log(P^{e}_{g|{\cal Z}}(n)) + 2\log(|a|+\delta) < 0,\]
\[\lim_{n \to \infty} (\kappa {1 \over n} \log(P^{e}_{g|g}(n)) + 2\log(|a|+\delta) + 2 \kappa \log(\alpha)< 0,\]
\[R'(n) > n\log_2(|a|/\alpha)\]
with arbitrarily small, positive $\eta > 0$ and
 \[ \kappa < {1 \over \log_{{|a|+\delta \over |a|}}({|a| + \delta \over \alpha})}.  \]
\qed \end{thm}

\textbf{Proof:} See Section \ref{ProofFiniteMoment}. \qed

Let us define
\[\bar{P}_e(n): = \max_{c \in {\cal M}(n)} P(\gamma(q'_{[0,n-1]}) \neq c | c \quad \mbox{is transmitted}).\]
When the block-length is clear from the context, we drop the index $n$. We have the following corollary to Theorem \ref{FiniteMomentChannel}.

\begin{cor}\label{FiniteMomentChannelDMCUniformErrorCase}
A sufficient condition for second moment stability (for the joint $(x_t,\Delta_t)$ process) over a discrete memoryless channel (DMC) is that:
\[\lim_{n \to \infty} (\kappa {1 \over n} \log(\bar{P}_e(n)) + 2\log(|a|+\delta) < 0,\]
with rate $R'(n) > n\log_2({|a|\over \alpha})$.
\qed \end{cor}

\begin{remark}\label{RandomCodeBound0}
For a DMC with block length $n$, Shannon's random coding \cite{GallagerIT85} satisfies:
\[P_e(n) \leq e^{-n E (R) + o(n)}, \]
uniformly for all codewords $c \in \{1,2,\dots, {\cal M}(n) \}$ with $c'$ being the decoder output (thus, the random exponent also applies uniformly over the set). Here ${o(n) \over n}\to 0$ as $n \to \infty$ and $E(R) > 0$ for $0 < R < C$. Thus, under the above conditions, the exponent under random coding should satisfy $E(R) > {2 \log_2(|a|+\delta) \over \kappa}$.
\end{remark}

\begin{remark}
The error exponent with feedback is typically improved with feedback, unlike capacity of DMCs. However, a precise solution to the error exponent problem of fixed length block coding with noiseless feedback is not known. Some partial results have been reported in \cite{Dobrushin} (in particular, the sphere packing bound is optimal for a class of symmetric channels for rates above a critical number even with feedback), Chapter 10 of \cite{CsiszarKorner}, \cite{BerlekampThesis}, \cite{Haroutunian}, \cite{Dyachkov}, \cite{Zigangirov}, \cite{Borade} and \cite{Nakiboglu2}. Particularly related to this section, \cite{Borade} has considered the exponent maximization for a special message symbol, at rates close to capacity. At the end of the paper, a discussion for variable length coding, in the context of Burnashev's \cite{Burnashev} setup, will be discussed along with some other contributions in the literature. In case feedback is not used, Gilbert exponent \cite{Omura} for low-rate regions may provide better bounds than the random coding exponent.
\qed
\end{remark}

\subsubsection{Zero-Error Transmission for ${\cal Z}$}

An important practical setup would be the case when ${\cal Z}$ is transmitted with no error and is not confused with messages from the granular messages. We state this as follows.

\textbf{Assumption A0}
We have that $P^e_{{\cal Z}|g}(n)=P^e_{g|{\cal Z}}(n)=0$ for $n \geq n_0$ for some $n_0 \in \mathbb{N}$.
 \qed

%
%
%

\begin{thm}\label{FiniteMomentChannelA1}
Under Assumption {\bf A0}, a sufficient condition for second moment stability is:
\[\lim_{n \to \infty} (\bar{P}_e(n)) (|a|+\delta)^{2n} < 1, \]
with rate $R'(n) > n\log_2({|a|\over \alpha})$ and $\kappa > 1/2$.
\qed \end{thm}


\textbf{Proof:} See Section \ref{ProofFiniteMomentChannelA1}. \qed

%

%
%
%

\begin{remark}
The result under ({\bf A0}) is related to the notion of any-time capacity proposed by Sahai and Mitter \cite{SahaiParts}. We note that Sahai and Mitter considered also a block-coding setup, for the case when the noise is bounded, and were able to obtain a similar rate/reliability criterion as above. It is worth emphasizing that, the reliability for sending one symbol ${\cal Z}$ for the under-zoom phase allows an improvement in the reliability requirements drastically.
\qed
\end{remark}

\section{Channels with Memory}


\begin{defn}
Let {\bf Class A} be the set of channels which satisfy the following two properties: \\
a) The following Markov chain condition holds:
 \[q'_t \leftrightarrow q_t, q_{[0,t-1]}, q'_{[0,t-1]} \leftrightarrow x_{[0,t]},\]
for all $t \geq 0$.\\
b) The channel capacity with feedback is given by:
\begin{eqnarray}\label{feedbackCapacity}
 C= \lim_{T \to \infty} \max_{\{ P(q_t | q_{[0,t-1]},q'_{[0,t-1]})\}} {1 \over T} I(q_{[0,T-1]} \to q'_{[0,T-1]}),
\end{eqnarray}
where $0 \leq t \leq T-1$ and the directed mutual information is defined by
\begin{eqnarray*}
&&I(q_{[0,T-1]} \to q'_{[0,T-1]}) \\
&& \quad \quad := \sum_{t=1}^{T-1} I(q_{[0,t]};q'_t|q'_{[0,t-1]}) + I(q_0;q'_0).
\end{eqnarray*}
\end{defn}

Discrete memoryless channels (DMCs) naturally belong to this class of channels. For such channels, it is known that feedback does not increase the capacity. Such a class also includes finite state stationary Markov channels which are indecomposable \cite{PermuterWeissmanGoldsmith}, and non-Markov channels which satisfies certain symmetry properties \cite{SenAlaYukIT}. Further examples are studied in \cite{TatikondaIT} and in \cite{DaboraGoldsmith}.

\begin{thm}\label{NecessityRecurrence2}
For a linear system controlled over a noisy channel with memory with feedback in {\bf Class A}, if the channel capacity is less than $\log_{2}(|a|)$, then the AMS property under the following condition
\[\liminf_{T \to \infty} {1 \over T} h(x_T) \leq 0, \]
cannot be satisfied under any policy.
\qed \end{thm}

\textbf{Proof:}
See the proof of Theorem \ref{StochasticStabilityofVectorNecessity} in Section \ref{proofNecessity}. \qed

The proof of the above is presented in Section \ref{proofNecessity}$^1$\footnotetext[1]{One can also obtain a positive result: If the channel capacity is greater than $\log_2(|a|)$ then there exists a coding scheme leading to an AMS state process provided that the channel restarts itself with the sending of a new block. If this assumption does not hold, then, using the proofs in the paper we can prove coordinate-recurrence under this condition. For the AMS property, however, new tools will be required.  Our proof would have to be modified to account for the non-Markovian nature of the sampled state and quantizer process.}. If the channel is not information stable, then information spectrum methods leads to pessimistic realizations of capacity (known as the $\liminf$ in probability of the normalized information density, see \cite{VerduHan}, \cite{TatikondaIT}). We do not consider such channels in this paper, although the proof is generalizable to some cases when the channel state is Markov and the worst case initial input state is considered as in \cite{PermuterWeissmanGoldsmith}. 

\section{Higher Order Sources}\label{MultiDimensionalSection}

The proposed technique is also applicable to a class of settings for the multi-dimensional setup. Observe that a higher order ARMA model of the form (\ref{ARMA}) can be written in the following form:
\begin{eqnarray}\label{vectorEqn1}
x_{t+1}=Ax_t + Bu_t + G d_t,
\end{eqnarray}
where $x_t \in \mathbb{R}^N$ is the state at time $t$, $u_t$ is the control input, and $\{d_t \}$ is a sequence of zero-mean independent, identically distributed (i.i.d.) zero-mean Gaussian random vectors of appropriate dimensions. Here $A$ is the system matrix with at least one eigenvalue greater than $1$ in magnitude, that is, the system is open-loop unstable. Furthermore, $(A,B)$ and $(A,G)$ are controllable pairs, that is, the state process can be traced in finite time from any point in $\mathbb{R}^N$ to any other point in at most $N$ time stages, by either the controller, or the Gaussian noise process.

In the following we assume that all modes with eigenvalues $\{\lambda_i, 1 \leq i \leq n\}$ of $A$ are unstable, that is have magnitudes greater than or equal to 1. There is no loss here since if some eigenvalues are stable, by a similarity transformation, the unstable modes can be decoupled from stable modes; stable modes are already recurrent.

\begin{thm}\label{StochasticStabilityofVectorNecessity}
Consider a multi-dimensional linear system with unstable eigenvalues, that is $|\lambda_i|\geq 1$ for $i=1,\dots,N$.
For such a system controlled over a noisy channel with memory with feedback in {\bf Class A}, if the channel capacity satisfies
\[C < \sum_{i} \log_2(|\lambda_i|),\]
there does not exist a stabilizing coding and control scheme with the property $\liminf_{T \to \infty} {1 \over T} {h(x_T)} \leq 0$.
\qed \end{thm}
\textbf{Proof:}
See Section \ref{proofNecessity}.  \qed

For sufficiency, we will assume that $A$ is a diagonalizable matrix with real eigenvalues (a sufficient condition being that the poles of the system are distinct real). In this case, the analysis follows from the discussion for scalar systems; as the identical recurrence analysis for the scalar case is applicable for each of the subsystems along each of the eigenvectors. The possibly correlated noise components will lead to the recurrence analysis discussed earlier. We thus have the following result.

\begin{thm}\label{StochasticStabilityofVector}
Consider a multi-dimensional system with a diagonalizable matrix $A$.
 If the Shannon capacity of the (DMC) channel used in the controlled system satisfies
\[C > \sum_{|\lambda_i|>1} \log_2(|\lambda_i|),\]
there exists a stabilizing (in the AMS sense) scheme.
\qed \end{thm}

\textbf{Proof:}
See Section \ref{ProofForVectorCaseAMS}.  \qed

The result can be extended to a case where the matrix $A$ is in a Jordan form. Such an extension entails considerable details in the analysis for the stopping time computations and has not been included in the paper. A discussion for the special case of discrete noiseless channels is contained in \cite{AndrewJohnstonReport} in the context of decentralized linear systems.

\section{Proofs}

\subsection{Proof of Theorem \ref{StochasticStabilityofVectorNecessity}}\label{proofNecessity}
We present the proof for a multi-dimensional system since this case is more general. For channels under {\bf Class A} (which includes the discrete memoryless channels (DMC) as a special case), the capacity is given by (\ref{feedbackCapacity}).

Let us define
\begin{eqnarray*}
R_T := && \max_{\{ P(q_t | q_{[0,t-1]},q'_{[0,t-1]}), 0 \leq t \leq T-1\}} \\
&& \quad \quad \quad {1 \over T}\sum_{t=0}^{T-1} I(q'_t; q_{[0,t]} | q'_{[0,t-1]})
\end{eqnarray*}
Observe that for $t > 0$:
\begin{eqnarray}
&& I(q'_t; q_{[0,t]} | q'_{[0,t-1]}) \nonumber \\
&& = H(q'_t | q'_{[0,t-1]}) -  H(q'_t | q_{[0,t]} ,q'_{[0,t-1]}) \nonumber \\
&& = H(q'_t | q'_{[0,t-1]}) -  H(q'_t | q_{[0,t]}, x_t,q'_{[0,t-1]}) \label{classAassumption} \\
&& \geq H(q'_t | q'_{[0,t-1]}) -  H(q'_t | x_t,q'_{[0,t-1]}) \nonumber \\
&& = I(x_t; q'_t| q'_{[0,t-1]}).
\end{eqnarray}
Here, (\ref{classAassumption}) follows from the assumption that the channel is of {\bf Class A}. It follows that since for two sequences such that $a_n \geq b_n$: $\limsup_n a_n \geq \limsup_n b_n$ and $R_T$ is assumed to have a limit:
\begin{eqnarray} \label{bounde}
&& \lim_{T \to \infty} R_T \nonumber \\
&&\geq  \limsup_{T \to \infty} {1 \over T} \bigg( \sum_{t=1}^{T-1} I(x_t; q'_t| q'_{[0,t-1]})) + I(x_0;q'_0) \bigg)\nonumber \\
 &&=  \limsup_{T \to \infty} {1 \over T}\bigg( \sum_{t=1}^{T-1} \bigg(h(x_t | q'_{[0,t-1]}) - h(x_t|q'_{[0,t]})\bigg) \nonumber \\
 && \quad \quad \quad\quad \quad \quad \quad + I(x_0;q'_0)\bigg)\nonumber \\
 &&= \limsup_{T \to \infty} {1 \over T}\bigg( \sum_{t=1}^{T-1} \bigg(h(A x_{t-1} + Gd_{t-1} + Bu_{t-1} |q'_{[0,t-1]} )\nonumber \\
 && \quad\quad \quad\quad \quad \quad \quad - h(x_t | q'_{[0,t]})\bigg) +I(x_0;q'_0)\bigg)\nonumber \\
 &&= \limsup_{T \to \infty} {1 \over T}\bigg( \sum_{t=1}^{T-1} \bigg(h(A x_{t-1} + Gd_{t-1} |q'_{[0,t-1]} ) \nonumber \\
 && \quad\quad \quad\quad \quad \quad \quad - h(x_t | q'_{[0,t]})\bigg) + I(x_0;q'_0)\bigg)\label{controlFunctionofPast} \\
 &&\geq \limsup_{T \to \infty} {1 \over T}\bigg( \sum_{t=1}^{T-1} \bigg(h(A x_{t-1} + Gd_{t-1} |q'_{[0,t-1]},d_{t-1} )\nonumber \\
 && \quad\quad \quad\quad \quad \quad \quad - h(x_t | q'_{[0,t]})\bigg) +I(x_0;q'_0)\bigg)\label{Abhishek} \\
 &&= \limsup_{T \to \infty} {1 \over T}\bigg( \sum_{t=1}^{T-1} \bigg( h(A x_{t-1}|q'_{[0,t-1]},d_{t-1} ) \nonumber \\
 && \quad\quad \quad \quad \quad \quad \quad- h(x_t | q'_{[0,t]}) \bigg) + I(x_0;q'_0)\bigg)\nonumber
 \end{eqnarray}
 \begin{eqnarray}
 &&= \limsup_{T \to \infty} {1 \over T}\bigg( \sum_{t=1}^{T-1} \bigg(h(A x_{t-1}|q'_{[0,t-1]}) \nonumber \\
 && \quad\quad \quad \quad \quad \quad \quad - h(x_t | q'_{[0,t]})\bigg) +I(x_0;q'_0)\bigg)\label{urbana} \\
 &&= \limsup_{T \to \infty} {1 \over T}\bigg( \sum_{t=1}^{T-1} \bigg(\log_2(|A|)+ h(x_{t-1}|q'_{[0,t-1]}) \nonumber \\
 && \quad\quad \quad \quad \quad \quad \quad - h(x_t | q'_{[0,t]})\bigg) + I(x_0;q'_0)\bigg)\nonumber \\
 &&= \limsup_{T \to \infty} {1 \over T}\bigg( \bigg(\sum_{t=1}^{T-1} \log_2(|A|) \bigg) + h(x_0|q'_0) \nonumber \\
 && \quad\quad \quad \quad \quad \quad \quad - h(x_{T-1} | q'_{[0,T-1]}) + I(x_0;q'_0)\bigg) \nonumber \\
 &&=  \log_2(|A|) - \liminf_{T \to \infty} \bigg( {1 \over T} h(x_{T-1} | q'_{[0,T-1]}) \bigg)  \nonumber \\
&&\geq  \log_2(|A|) - \liminf_{T \to \infty} \bigg( {1 \over T} h(x_{T-1}) \bigg)  \label{conditionagain} \\
&&\geq \log_2(|A|). \nonumber
\end{eqnarray}
 Equation (\ref{controlFunctionofPast}) follows from the fact that the control action is a function of the past channel outputs, (\ref{Abhishek}) follows from the fact that conditioning does not increase entropy, and (\ref{urbana}) from the observation that $\{d_t\}$ is an independent process. (\ref{conditionagain}) follows from conditioning. The other equations follow from the properties of mutual information. By the hypothesis, $\liminf_{t \to \infty} {1 \over t} h(x_t) \leq 0$,
it must be that $\lim_{T \to \infty} R_T \geq \log_2(|A|)$. Thus, the capacity also needs to satisfy this bound. \qed

\subsection{Stopping Time Analysis}\label{ProofStocStabSystemTimeDistribution}

This section presents an important supporting result on stopping time distributions, which is key in the application of Theorem \ref{thm5} for the stochastic stability results. We begin with the following.

\begin{lem}\label{MarkovChain}
Let ${\cal B}(\mathbb{R} \times {\mathbb{R}_+})$ denote the Borel $\sigma-$field on $\mathbb{R} \times {\mathbb{R}_+}$. It follows that
\begin{eqnarray}
&&P\bigg((x_{kn},\Delta_{kn}) \in (C\times D) | \{(x_{sn},\Delta_{sn}), s < k\} \bigg) \nonumber \\
&& = P\bigg((x_{kn},\Delta_{kn}) \in (C\times D)| (x_{(k-1)n},\Delta_{(k-1)n})\bigg), \nonumber
\end{eqnarray}
 $\forall (C\times D) \in {\cal B}(\mathbb{R} \times {\mathbb{R}_+}),$ i.e. $(x_{tn},\Delta_{tn})$ is a Markov chain.
\qed \end{lem}

The above follows from the observations that, the channel is memoryless, the encoding only depends on the most recent samples of the state and the quantizer, and the control policies use the channel outputs received in the last block, which stochastically only depend on the parameters in the previous block.


Let us define $h_t:= { x_t \over \Delta_t 2^{R'-1}}$. We will say that the quantizer is perfectly zoomed when $|h_t| \leq 1$, and under-zoomed otherwise.

Define a sequence of stopping times for the perfect-zoom case with (where the initial state is perfectly zoomed at $\tau_0$)
\begin{eqnarray}\label{stopTimes}
\tau_{0} = 0, \quad \tau_{z+1} = \inf \{kn > \tau_z : |h_{kn}| \leq  1 \}, \quad z, k \in \mathbb{Z}_+ 
\end{eqnarray}

As discussed in Section \ref{sectionFiniteSecondMoment}, there will be three types of errors.
\begin{itemize}
\item Type I-A: Error from a granular symbol to another granular symbol. In this case the quantizer will zoom in, yet an incorrect control action will be applied. As in Section \ref{sectionFiniteSecondMoment}, $P^e_{g|g}(n)$ is an upper bound for such an error.
\item Type I-B: Error from a granular symbol to ${\cal Z}$. In this case, no control action will be applied and the quantizer will be zoomed out. As in Section \ref{sectionFiniteSecondMoment}, $P^e_{{\cal Z}|g}(n)$ is an upper bound for such an error.
\item Type II: Error from ${\cal Z}$ to a granular symbol. At consecutive time stages, until the next stopping time, the quantizer should ideally zoom out. Hence, this error may take place in subsequent time stages (since at time $0$ the quantizer is zoomed, this does not take place). The consequence of such an error is that the quantizer will be zoomed in and an incorrect control action will be applied. Let
\begin{eqnarray}
&& P_e(n):= P^e_{g|{\cal Z}}(n)  \nonumber \\
&& \quad = P(\gamma(q'_{[0,n-1]}) \neq {\cal Z} | {\cal Z} \quad \mbox{is transmitted}) \nonumber
\end{eqnarray}
\end{itemize}
We will drop the dependency on $n$, when the block length is clear from the context.


\begin{lem}\label{keyLemma}
The discrete probability distribution $\Prob(\tau_{z+1}-\tau_z|x_{\tau_z},\Delta_{\tau_z})$ is asymptotically, in the limit of large $\Delta_{\tau_z }$, dominated (majorized) by a geometrically distributed measure. That is, for $k \geq \lceil 1/\kappa \rceil + 1$,
\begin{eqnarray}
&&\Prob(\tau_{z+1}-\tau_{z} \geq kn | x_{\tau_{z}},\Delta_{\tau_{z}}) \nonumber \\
&& \leq
 \Xi(\Delta_{\tau_z}) \bigg( (1- P^e_{g|g} - P^e_{{\cal Z}|g}) (eP_e^{(\kappa)})^{k-2} \nonumber \\
&& \quad \quad \quad \quad \quad \quad + P^e_{g|g} (eP_e^{(\kappa - {1 - \kappa \over k-2})})^{k-2} \nonumber \\
&& \quad \quad \quad \quad \quad \quad + (P^e_{{\cal Z}|g}) (eP_e^{(\kappa + { \kappa \over k-2})})^{k-2} \bigg)
\end{eqnarray}
 where $\Xi(\Delta_{\tau_z}) < \infty$ and $\Xi(\Delta_{\tau_z}) \to 1$ as $\Delta_{\tau_z} \to \infty$ for every fixed $n$, uniformly in $|h_0| \leq 1$ and
\begin{eqnarray}\label{kappaDefinitionOld}
\kappa < {1 \over \log_{{|a|+\delta \over |a|}}({|a| + \delta \over \alpha})}.
\end{eqnarray}
 \qed
\end{lem}

\textbf{Proof:} \\
Denote for $k \in \mathbb{N}$,
 \begin{equation}
\Theta_k := \Prob(\tau_{z+1} - \tau_{z} \geq kn|x_{\tau_{z}},\Delta_{\tau_{z}}).
\label{e:stpP}
\end{equation}
Without any loss, let $z=0$ and $\tau_{0}=0$, so that $\Theta_k = \Prob(\tau_{1} \geq kn |x_0,\Delta_{0}) $.

Before proceeding with the proof, we highlight the technical difficulties that will arise when the quantizer is in the under-zoom phase. As elaborated on above, the errors at time $0$ are crucial for obtaining the error bounds: At time $0$, at most with probability $P^e_{g|g}(n)$, an error will take place so that the quantizer will be zoomed in, yet an incorrect control signal will be applied. With probability at most $P^e_{{\cal Z}|g}(n)$, an error will take place so that, no control action is applied and the quantizer is zoomed out. At consecutive time stages, until the next stopping time, the quantizer should ideally zoom out but an error takes place with probability $P^e_{g|{\cal Z}}(n)$ and leads the quantizer to be zoomed in, and a control action to be applied. Our analysis below will address all of these issues.

In the following we will assume that the probabilities are conditioned on particular $x_0,\Delta_0$ values, to ease the notational presentation.

We first consider the case where there is an intra-granular, Type I-A, error at time $0$, which takes place at most with probability $P^e_{g|g}$ (this happens to be the worst case error for the stopping time distribution). Now,
\begin{eqnarray}\label{startTheBound}
&& \Prob(\tau_{1} \geq kn | \mbox{Type I-A error at time $0$}) \nonumber \\
&& = \Prob\bigg( \bigcap_{m=1}^{k-1} (|h_{mn}| > 1) | \mbox{Type I-A error at time $0$} \bigg) \nonumber \\
&&= \Prob \bigg( \bigcap_{m=1}^{k-1} (|x_{mn}| \geq 2^{R'-1} (|a|+\delta)^{(m-s_m-1)n} \nonumber \\
&& \quad \quad \quad \quad \quad \quad \times \alpha^{(1+s_m)n}\Delta_0)   \bigg) \nonumber \\
&&= \Prob \bigg( \bigcap_{m=1}^{k-1} (|a^{mn}(x_0 + \sum_{i=0}^{mn-1}a^{-i-1}(d_i + u_i))| \nonumber \\
&& \quad \quad \quad \quad \geq 2^{R'-1} (|a|+\delta)^{(m-s_m-1)n}\alpha^{(1+s_m)n}\Delta_0)   \bigg) \nonumber \\
&&= \Prob \bigg( \bigcap_{m=1}^{k-1} (|(x_0 + \sum_{i=0}^{mn-1}a^{-i-1}(d_i + u_i))| \nonumber \\
&& \quad \quad  \geq {2^{R'-1} \alpha^n \over |a^n|} ({|a|+\delta \over |a| })^{(m-1)n}({\alpha \over |a|+\delta})^{(s_m)n}\Delta_0)   \bigg) \nonumber \\
\end{eqnarray}

In the above, $s_m$ is the number of errors in the transmissions that have taken place until (but not including) time $m$, except for the one at time $0$. An error at time $0$ would possibly lead to a further enlargement of the bin size with non-zero probability; whereas no-error at time 0 leads to a strict decrease in the bin size.

The study for the number of errors is crucial for analyzing the stopping time distributions. In the following, we will condition on the number of erroneous transmissions for $k$ successive block codings for the under-zoomed phase. Suppose that for $k>1$, there are $s_{k}$ total erroneous transmissions in the time stages $\{n, 2n, \dots, (k-1)n\}$ when the state is in fact under-zoomed, but the controller interprets the received transmission as a successful one. Thus, we take $s_1=0$.

Let $\zeta_1, \zeta_2, \dots, \zeta_{s_{k-1}}$ be the time stages when errors take place, such that
\[\zeta_{t+1} : \min(\min(m > \zeta_{t}: c'_{nm} \neq c_{nm}) , k-1), \quad \zeta_0 = 0,\]
such that $\zeta_{s_{k-1}+1}=k-1$ or $\zeta_{s_{k-1}}=k-1$ and define $\eta_t = \zeta_{t+1} - \zeta_t$.

In the time interval $[\zeta_{t}n+1, \zeta_{t+1}n-1]$ the system is open-loop, that is there is no control signal applied, as there is no misinterpretation by the controller. However, there will be a non-zero control signal at times $\{\zeta_{k}n, k \geq 0\}$. These are, however, upper bounded by the ranges of the quantizers at the corresponding time stages. That is, when an erroneous interpretation by the controller arises, the control applied $-(a^n/b) u_{(\zeta_z+1)n-1}$ lives in the set: $\{a^n (- 2^{R'-1} + k- (1/2)) \Delta_{\zeta_z}, 1 \leq k \leq 2^{R'} \} $.

\begin{figure*}[!t]
\normalsize
\setcounter{mytempeqncnt}{\value{equation}}
\setcounter{equation}{19}
\begin{eqnarray}
&& \Prob\bigg( \bigcap_{m=1}^{k-1} (|h_{mn}| > 1) | \mbox{Type I-A error at time $0$} \bigg)\nonumber \\
&& \leq \Prob \bigg( \bigcap_{m=1}^{k-1} \bigg( |a^{mn}(x_0 + \sum_{i=0}^{mn-1}a^{-i-1}(d_i + u_i))| \geq 2^{R'-1} (|a|+\delta)^{(m-s_m-1)n}\alpha^{(1+s_m)n}\Delta_0 \bigg)   \bigg) \nonumber \\
&& \leq \Prob \bigg(\bigcup_{p=0}^{k-2} \bigg( \{s_{k-1}=p\} \bigcap \bigg\{ \bigcap_{m=1}^{p} (|a^{\zeta_m n}(x_0 + \sum_{i=0}^{\zeta_mn-1} a^{-i-1}d_i + \sum_{i=0}^{m-1} a^{(- \zeta_i -1)n} u_{(\zeta_i+1)n-1}) |   \nonumber \\
&& \quad \quad \quad \quad \quad \quad \quad \quad \quad \quad  \geq 2^{R'-1} (|a|+\delta)^{(\zeta_m- s_m - 1)n}\alpha^{(1+s_m)n} \Delta_{0}) \bigg\} \bigg)  \bigg) \nonumber \\
&&= \sum_{p=0}^{k-2} {k-2 \choose p} (P^e_{g|{\cal Z}})^{p} (1 - (P^e_{g|{\cal Z}}))^{k-1-p} 1_{\{s_{k-1}=p\}} P\bigg( \bigcap_{m=1}^{p} (|a^{\zeta_m n}(x_0 + \sum_{i=0}^{\zeta_mn-1} a^{-i-1}d_i + \sum_{i=0}^{m-1} a^{(- \zeta_i -1)n} u_{(\zeta_i+1)n-1}) |   \nonumber \\
&& \quad \quad \quad  \quad \quad \quad \quad \quad  \geq 2^{R'-1} (|a|+\delta)^{(\zeta_m- s_m - 1)n}\alpha^{(1+s_m)n} \Delta_0 \bigg| s_{k-1}=p \bigg) \label{addedLater}
\end{eqnarray}
\setcounter{equation}{\value{mytempeqncnt}}
\hrulefill
\vspace*{4pt}
\end{figure*}
\addtocounter{equation}{1}

From (\ref{startTheBound}), we obtain (\ref{addedLater}). Regarding (\ref{addedLater}), let us now observe the following:
\begin{eqnarray}
&&P(\bigg|a^{\zeta_m n}(x_0 + \sum_{i=0}^{\zeta_mn-1} a^{-i-1}d_i \nonumber \\
&& \quad \quad \quad + \sum_{i=0}^{m-1} a^{(- \zeta_i -1)n} u_{(\zeta_i+1)n-1}) \bigg| \nonumber \\
&& \quad \quad \quad \quad \geq 2^{R'-1} (|a|+\delta)^{(\zeta_m- s_m - 1)n}\alpha^{(1+s_m)n} \Delta_0  \bigg)  \nonumber \\
&&\leq P\bigg(|\sum_{i=0}^{\zeta_mn-1} a^{-i-1}d_i|  \nonumber \\
&& \quad  \quad \quad \quad  \geq 2^{R'-1} ({|a|+\delta \over |a|})^{(\zeta_m- s_m - 1)n} ({\alpha \over |a|})^{(1+s_m)n} \Delta_0 \nonumber \\
&& \quad \quad \quad \quad \quad- |x_0 +  \sum_{i=0}^{m-1}  a^{(- \zeta_i -1)n} u_{(\zeta_i+1)n-1}|  \bigg)  \nonumber
\end{eqnarray}

\begin{figure*}[!t]
\normalsize
\setcounter{mytempeqncnt}{\value{equation}}
\setcounter{equation}{20}
\begin{eqnarray}
&& P\bigg\{ \bigcap_{m=1}^{p} \bigg(|a^{\zeta_m n}(x_0 + \sum_{i=0}^{\zeta_m n-1} a^{-i-1}d_i + \sum_{i=0}^{m-1} a^{(- \zeta_i -1)n} u_{(\zeta_i+1)n-1}) | \geq 2^{R'-1} (|a|+\delta)^{(\zeta_m- s_m-1)n}\alpha^{(1+s_m)n} \Delta_0 \mid s_{k-1}=p \bigg) \bigg\} \nonumber \\
&& \leq P\bigg\{ \bigcap_{m=1}^{p} \bigg(|\sum_{i=0}^{\zeta_mn-1} a^{-i-1}d_i| \geq 2^{R'-1} {a^{-\zeta_m n}} (|a|+\delta)^{(\zeta_m- s_m-1)n}\alpha^{(1+s_m)n}\Delta_0 \nonumber \\
&& \quad \quad \quad \quad \quad \quad \quad \quad \quad \quad \quad \quad -|x_0| - |\sum_{i=0}^{m-1} a^{(- \zeta_i -1)n} u_{(\zeta_i+1)n-1}|  \bigg| s_{k-1}=p \bigg) \bigg\} \nonumber \\
&& \leq \min_{0 \leq m \leq s_{k-1}} \bigg\{ P\bigg(|\sum_{i=0}^{\zeta_mn-1} a^{-i-1}d_i|  \geq 2^{R'-1} ({|a|+\delta \over |a|})^{(\zeta_m- s_m-1)n} ({\alpha \over |a|})^{(1+s_m)n} \Delta_0 - |x_0| - (2^{R'-1} - 1/2) \Delta_0 \nonumber \\
&& \quad \quad \quad \quad \quad \quad \quad \quad \quad \quad - \sum_{i=1}^{m-1} |a|^n ({|a|+\delta \over |a|})^{(\zeta_i - s_i-1)n} ({\alpha \over |a|})^{(s_m+1) n} (2^{R'-1} - 1/2) \Delta_0 \bigg| s_{k-1}=p \bigg) \bigg\} \label{x0LessThanTheEntireRange} \\
&& \leq \min_{0 \leq m \leq s_{k-1}} \bigg\{ P\bigg(|\bar{d}| \geq 2^{R'-1} ({|a|+\delta \over |a|})^{(\zeta_m- s_m-1)n} ({\alpha \over |a|})^{(1+s_m)n} \Delta_0 - |x_0| - (2^{R'-1} - 1/2) \Delta_0 \nonumber \\
&& \quad \quad \quad \quad \quad \quad \quad \quad -  \sum_{i=1}^{m-1}  ({|a|+\delta \over |a|})^{(\zeta_m - s_m-1)n} ({\alpha \over |a|})^{(1+s_m)n} (2^{R'-1} - 1/2) \Delta_0 \bigg| s_{k-1}=p \bigg) \bigg\} \label{ohio0}
\end{eqnarray}
\setcounter{equation}{\value{mytempeqncnt}}
\hrulefill
\vspace*{4pt}
\end{figure*}
\addtocounter{equation}{2}
Since the control signal $u_{(\zeta_i+1)n-1}$ lives in: $\{a^n (- 2^{R'-1} + k- (1/2)) \Delta_{(\zeta_i)n}, 1 \leq k \leq 2^{R'} \}$, conditioned on having $s_{k-1}$ errors in the transmissions, the bound writes as (\ref{ohio0}), where $\bar{d} = \sum_{i=0}^{\infty} a^{-i-1}d_i$ is a zero-mean Gaussian random variable with variance ${E[d^2]a^{-2} \over 1- a^{-2}}$. Here, (\ref{x0LessThanTheEntireRange}) considers the worst case when even when the quantizer is zoomed, the controller incorrectly picks the worst case control signal and the chain rule for total probability: For two events $A,B$: $P(A,B) \leq \min(P(A),P(B))$. The last inequality follows since $P(|\bar{d}| \geq \alpha_B) \geq P(|\sum_{i=0}^{\zeta_mn-1} a^{-i-1}d_i| \geq \alpha_B)$ for any $\alpha_B \in \mathbb{R}$.

\begin{figure*}[!t]
\normalsize
\setcounter{mytempeqncnt}{\value{equation}}
\setcounter{equation}{22}
\begin{eqnarray}\label{analysisSumBound}
&& P \bigg( |\bar{d}|  \geq 2^{R'-1} ({|a|+\delta \over |a|})^{(\zeta_m- s_m-1)n} ({\alpha \over |a|})^{(1+s_m)n} - |x_0| - (2^{R'-1} - 1/2) \Delta_0 \nonumber \\
&& \quad \quad \quad \quad \quad \quad \quad \quad \quad \quad -  \sum_{i=1}^{m-1}  ({|a|+\delta \over |a|})^{(\zeta_m - s_m-1)n} ({\alpha \over |a|})^{(1+s_m)n} (2^{R'-1} - 1/2) \Delta_0 \bigg| s_{k-1}=p \bigg) \nonumber \\
&& = P \bigg( |\bar{d}| \geq 2^{R'-1} ({|a|+\delta \over |a|})^{(\zeta_m - s_m)n} ({\alpha \over |a|})^{(s_m)n} ({\alpha \over |a| + \delta})^n \bigg(1 -  \sum_{i=1}^{m-1}  ({|a|+\delta \over |a|})^{(\zeta_i - \zeta_m)n} ({\alpha \over |a|})^{(s_i-s_m)n} (1 - 2^{-R'}) \bigg) \Delta_0 \nonumber \\
&& \quad \quad \quad \quad \quad \quad \quad \quad \quad \quad - |x_0| - (2^{R'-1} - 1/2)\Delta_0 \bigg| s_{k-1}=p \bigg) \nonumber \\
&& \leq P \bigg( |\bar{d}| \geq 2^{R'-1} ({|a|+\delta \over |a|})^{(\zeta_m - s_m)n} ({\alpha \over |a|})^{(s_m)n} ({\alpha \over |a| + \delta})^n \bigg(1 -  \sum_{i=1}^{m-1}  ({|a|+\delta \over |a|})^{(\zeta_i - \zeta_m)n} ({\alpha \over |a|})^{(s_i-s_m)n} (1 - 2^{-R'}) \bigg) \Delta_0 \nonumber \\
&& \quad \quad\quad \quad \quad \quad \quad  \quad \quad \quad \quad \quad \quad \quad \quad \quad \quad \quad - 2^{R'}\Delta_0 \bigg| s_{k-1}=p \bigg)
\end{eqnarray}
\setcounter{equation}{\value{mytempeqncnt}}
\hrulefill
\vspace*{4pt}
\end{figure*}
\addtocounter{equation}{1}

Now, let us consider $m=k-1$. In this case, (\ref{analysisSumBound}) follows, where in the last inequality we observe that $|x_0| \leq 2^{R'-1}\Delta_0$, since the state is zoomed at this time.

In bounding the stopping time distribution, we will consider the condition that
\begin{eqnarray}\label{newDefinitionKappa}
(k-1) - (s_{k-1}+1)(\log_{{(|a|+\delta) \over |a|}}({|a| + \delta \over \alpha })) > \alpha_A (s_{k-1} + 1)
\end{eqnarray}
 for some arbitrarily small but positive $\alpha_A$, to be able to establish that
\begin{eqnarray}\label{boundustunebound}
\bigg(1 -  \sum_{i=1}^{m-1}  ({|a|+\delta \over |a|})^{(\zeta_i - \zeta_m)n} ({\alpha \over |a|})^{(s_i-s_m)n} (1 - 2^{-R'}) \bigg) > 0
\end{eqnarray}
and that $({|a|+\delta \over |a|})^{(\zeta_{k-1} - s_{k-1})n} ({\alpha \over |a| + \delta})^{(s_{k-1})n} > 2$ for sufficiently large $n$. Now, there exists an $m$ such that
\begin{eqnarray}
&& ({|a|+\delta \over |a|})^{\zeta_m n} ({\alpha \over |a + \delta|})^{ s_m n} \nonumber \\
&& \quad \quad \quad \quad \geq ({|a|+\delta \over |a|})^{(\zeta_{k-1} - s_{k-1})n} ({\alpha \over |a| + \delta})^{(s_{k-1})n} \nonumber
\end{eqnarray}
 and for this $m$,
\begin{eqnarray}\label{uniformBound222}
 &&\bigg(1 -  \sum_{i=1}^{m-1}  ({|a|+\delta \over |a|})^{n(\zeta_i - \zeta_m - (\log_{{(|a|+\delta) \over |a|}}({|a| + \delta \over \alpha }))(s_i-s_m))} \bigg) \nonumber \\
 && \quad \quad \quad \quad \quad \quad \geq (1 - \sum_{i=1}^{m-1} ({|a|+\delta \over |a|})^{-\alpha_A i n}) > 0.
 \end{eqnarray}
This follows from considering a conservative configuration among an increasing subsequence of times $\{\zeta_i, \dots, \zeta_m\}$, such that for all elements of this sequence:
\begin{eqnarray*}
&&({|a|+\delta \over |a|})^{\zeta_i n} ({\alpha \over |a + \delta|})^{ s_i n} \\
&& \quad \quad \quad \geq ({|a|+\delta \over |a|})^{(\zeta_{k-1} - s_{k-1})n} ({\alpha \over |a| + \delta})^{(s_{k-1})n}
\end{eqnarray*}
and for every element up until time $m$,
$(\zeta_m - \zeta_{i} - (s_m-s_i)(\log_{{(|a|+\delta) \over |a|}}({|a| + \delta \over \alpha })) \geq \alpha_A(s_m-s_i)$. Such an ordered sequence provides a conservative configuration which yet satisfies (\ref{uniformBound222}), by considering if needed, $m$ to be an element in the sequence with a lower index value for which this is satisfied. This has to hold at least for one time $\zeta_m$, since $k-1$ satisfies (\ref{newDefinitionKappa}). Such a construction ensures that $(\ref{boundustunebound})$ is uniformly bounded from below for every $k$ since $\sum_{i=1}^{\infty} ({|a|+\delta \over |a|})^{-\alpha_A i n} < 1$ for $n$ large enough.

Hence, by (\ref{newDefinitionKappa}), for some constant $B_b > 0$, the following holds:
\begin{eqnarray}\label{ilkBound}
&& P\bigg( |\bar{d}| \geq  B_b \Delta_{0} \bigg( ({|a|+\delta \over a})^{\zeta_{m}n} ({\alpha \over (|a|+\delta)})^{(s_m+1)n} \bigg) \nonumber \\
&& \leq 2 {\sigma' \over B_b \Delta_{0}  \bigg( ({|a|+\delta \over a})^{\zeta_{mn}} ({\alpha \over |a|+\delta})^{(s_m+1)} \bigg) } \nonumber \\
&& \quad \quad \quad \times e^{- \bigg( B_b \Delta_{0}  ({|a|+\delta \over a})^{\zeta_{mn}} ({\alpha \over (|a|+\delta)})^{(s_m+1)n} \bigg)^2 / 2 \sigma'^2}
 \end{eqnarray}
The above follows from bounding the complementary error function by the following: $\int_q^{\infty} \mu(dx) \leq \int_q^{\infty} {x \over
q}\mu(dx)$, for $q>0$ when $\mu$ is a zero-mean Gaussian measure. In the above derivation $\sigma'^2 = E[d_1^2] |a|^{-2}/(1 - |a|^{-2})$. The left hand side of (\ref{ilkBound}) can be further upper bounded by, for any $r>0$:
 \begin{eqnarray}\label{boundProduct}
&& M_r(\Delta_0) r^{- \bigg(({|a|+\delta \over a})^{\zeta_{m}n} ({\alpha \over |a| +\delta})^{(s_m+1)n} \bigg)} \nonumber \\
&& \quad + \bigg( 1 - 1_{\{ \zeta_m - (s_{m}+1)(\log_{{(|a|+\delta) \over |a|}}({|a| + \delta \over \alpha })) > \alpha_A (s_{m} + 1)  \}} \bigg) \nonumber \\
\end{eqnarray}
with $M_r(\Delta_0) \to 0$ as $\Delta_0 \to \infty$ exponentially:
 \begin{eqnarray}\label{speedofMrTo0}
 {M_r(\Delta_0) \over \Delta_0^{-p}} \to 0,
 \end{eqnarray}
for any $p \in \mathbb{N}_+$, due to the exponential dependence of (\ref{ilkBound}) in $\Delta_0$.
Thus, combined with (\ref{newDefinitionKappa}), conditioned on having $s_{k-1}$ errors and a Type I-A error at time $0$, we have the following bound on (\ref{ohio0})
\begin{eqnarray}\label{star}
&& M_r(\Delta_0)  r^{- \bigg(({|a|+\delta \over a})^{(k-1-s_{k-1})n} ({\alpha \over |a|})^{(s_{k-1}+1)n} \bigg)} \nonumber \\
&& \quad \quad \quad \quad + 1_{\{ \zeta_{k-1} \leq {(s_{k-1}+1) \over \kappa}   \} }
\end{eqnarray}
with
\begin{eqnarray}\label{kappaDefinition}
\kappa = {1 \over \log_{{(|a|+\delta) \over |a|}}({|a| + \delta \over \alpha}) + \alpha_A}.
\end{eqnarray}

We observe that the number of errors needs to satisfy the following relation for the above bound in (\ref{boundProduct}) to be less than $1$:
\[k-1  > (1+s_{k-1}) / \kappa \]
Finally, the probability that the number of incorrect transmissions exceeds $\kappa(k-1)-1$ is exponentially low, as we observe in the following. Let, as before, $P_e(n)=P^e_{g|{\cal Z}}(n)$. We consider below Chernoff-Sanov's Theorem: The sum of Bernoulli error events leads to a binomial distribution. Let for $1>\zeta>0$, $D(\zeta,P_e) = \zeta\log(\zeta/P_e) + (1 - \zeta)\log({1-\zeta \over 1-P_e} )$. Then, the following upper bound holds \cite{Cover}, for every $k > 3$:
\begin{eqnarray}
&& P\bigg(  \sum_{t=1}^{k-2} 1_{\{\mbox{Type II Error}\}} \geq \kappa(k-1) - 1 \bigg) \nonumber \\
&& = P\bigg(  \sum_{t=1}^{k-2} 1_{\{\mbox{Type II Error}\}} \geq (k-2) (\kappa - {1 - \kappa \over k-2}) \bigg) \nonumber \\
&& \leq e^{-(k-2) D((\kappa - {1 - \kappa \over k-2}),P_e)} ,
\end{eqnarray}
Hence,
\begin{eqnarray}\label{ohio}
&&P\bigg(  \sum_{t=1}^{k-2} 1_{\{\mbox{Type II Error}\}} \geq (\kappa - {1 - \kappa \over k-2})(k-2) \bigg) \nonumber \\
&& \quad \quad \quad \quad \quad \quad \leq (e^{H((\kappa - {1 - \kappa \over k-2}))} P_e^{(\kappa - {1 - \kappa \over k-2})})^{k-2},
\end{eqnarray}
with $H(z) = -z \log(z) - (1 - z) \log(1 - z) \leq 1$. Hence,
\begin{eqnarray}\label{ohio00}
&& P\bigg(  \sum_{t=1}^{k-2} 1_{\{\mbox{Type II Error}\}} \geq (\kappa - {1 - \kappa \over k-2})(k-2) \bigg)  \nonumber \\
&& \quad \quad \quad \quad \quad \quad \leq (e P_e^{(\kappa - {1 - \kappa \over k-2})})^{k-2},
\end{eqnarray}

We could bound the following summation as follows.
\begin{eqnarray}
&& \sum_{s_{k-1}=0}^{\lfloor \kappa (k-1) \rfloor -1}  {k-2 \choose s_{k-1}} M_r(\Delta_0) r^{-\bigg( (k-1)  - (s_{k-1} + 1)/\kappa \bigg)n }  \nonumber \\
&& \quad \quad \quad \quad \quad \quad \times (P_e)^{s_{k-1}} (1-P_e)^{k-1-s_{k-1}} \label{tulsa} \\
&&\leq M_r(\Delta_0) (1-P_e)^{k-1} \bigg( \sum_{s_{k-1}=0}^{\lfloor (\kappa - {1 - \kappa \over k-2}) (k-2) \rfloor}  {k-2 \choose s_{k-1}} \bigg) \nonumber \\
&& \quad \quad \quad \quad \times ({P_e \over 1-P_e})^{\kappa(k-1)-1} \label{tulsa1} \\
&&\leq M_r(\Delta_0)  2^{(k-2)}  (P_e)^{(\kappa - {1 - \kappa \over k-2})(k-2)}  \nonumber \\
&&= M_r(\Delta_0)  (2P_e^{(\kappa - {1 - \kappa \over k-2})})^{(k-2)} \label{tulsa4}
\end{eqnarray}
where (\ref{tulsa})-(\ref{tulsa1}) holds since $r$ can be taken to be $r > ({1-P_e \over P_e})^{\kappa}$ by taking $\Delta_0$ to be large enough and in the summations $s_{k-1}$ taken to be $\kappa(k-1)-1$. We also use the inequality
\begin{eqnarray}\label{FlumGroheBound}
\sum_{s=0}^{\lfloor \kappa (k-1) \rfloor-1} {k-2 \choose s} \leq 2^{k-2},
\end{eqnarray}
and that $\kappa(k-1) - 1 \leq k-2$.


Thus, from (\ref{addedLater}) we have computed a bound on the stopping time distributions through (\ref{ohio00}) and (\ref{tulsa4}). Following similar steps for the Type I-B error and no error cases at time 0, we obtain the bounds on the stopping time distributions as follows:

\begin{itemize}
\item Conditioned an error in the granular region (Type I-A) at time $0$, the condition for the number of errors is that
\[k-1  > (1+s_{k-1}) {1 \over \kappa} \]
and by adding (\ref{ohio}) and (\ref{tulsa4}), the stopping time is dominated by:
\begin{eqnarray}\label{keyBoundSum1}
&& P(\tau_1 \geq kn) \nonumber \\
&& \quad \leq  M_r(\Delta_0)  (2P_e^{(\kappa - {1 - \kappa \over k-2})})^{(k-2)}  + (e P_e^{(\kappa - {1 - \kappa \over k-2})})^{k-2} \nonumber\\
&& \quad \leq \Xi(\Delta_0) (e P_e^{(\kappa - {1 - \kappa \over k-2})})^{k-2}
\end{eqnarray}
for $\Xi(\Delta) = M_r(\Delta) + 1$ which goes to 1 as $\Delta \to \infty$.

\item  Conditioned on the error that ${\cal Z}$ is the decoding output at time $0$, in the above, the condition for the number of errors is that
\[k-1  > s_{k-1} {1 \over \kappa} \]
and we may replace the exponent term $(\kappa - {1 - \kappa \over k-2})$ with $(\kappa + { \kappa \over k-2})$ and the stopping time is dominated by
\begin{eqnarray}\label{keyBoundSum2}
P(\tau_1 \geq kn) 
&\leq& \Xi(\Delta_0) (eP_e^{(\kappa + { \kappa \over k-2})})^{(k-2)}
\end{eqnarray}
for $\Xi(\Delta) = M_r(\Delta) + 1$ which goes to 1 as $\Delta \to \infty$.

\item  Conditioned on no error at time 0 and the rate condition $R' > \log_2(|a|/\alpha)$, the condition for the number of errors is that
\[k-1  > 1+ {s_{k-1} \over \kappa} \]
and we may replace the exponent term $(\kappa - {1 - \kappa \over k-2})$ with $\kappa$.

The reason for this is that, $|x_0-\hat{x}_0| \leq \Delta_0/2$ and the control term applied at time $n$ reduces the error.

\begin{figure*}[!t]
\normalsize
\setcounter{mytempeqncnt}{\value{equation}}
\setcounter{equation}{40}
\begin{eqnarray}\label{addedLater2}
&& P\bigg( \bigcap_{m=1}^{p} (|a^{\zeta_m n}(x_0 + \sum_{i=0}^{\zeta_mn-1} a^{-i-1}d_i + \sum_{i=0}^{m-1} a^{(- \zeta_i -1)n} u_{(\zeta_i+1)n-1}) | \nonumber \\
&& \quad \quad \quad \quad \quad \quad \quad \quad  \geq 2^{R'-1} (|a|+\delta)^{(\zeta_m- s_m - 1)n}\alpha^{(1+s_m)n}  | s_{k-1}=p \bigg) \nonumber \\
&& \leq \min_{0 \leq m \leq s_{k-1}} \bigg\{ P\bigg(|\sum_{i=0}^{\zeta_mn-1} a^{-i-1}d_i| \geq 2^{R'-1} ({|a|+\delta \over |a|})^{(\zeta_m- s_m - 1)n} ({\alpha \over |a|})^{(1+s_m)n} \nonumber \\
&& \quad \quad \quad \quad \quad \quad \quad \quad - |x_0 - \hat{x}_0| - \sum_{i=1}^{m-1} |a|^n ({|a|+\delta \over |a|})^{\zeta_i - s_i -1 } ({\alpha \over |a|})^{(s_i+1) n} (2^{R'-1} - 1/2) \Delta_0 \bigg) \bigg\} \nonumber \\
&& \leq \min_{0 \leq m \leq s_{k-1}} \bigg\{ P\bigg(|\bar{d}|  \geq (2^{R'-1} ({|a|+\delta \over |a|})^{(\zeta_m- s_m - 1)n} ({\alpha \over |a|})^{(1+s_m)n} - 1/2) \Delta_0 \nonumber \\
&& \quad \quad \quad \quad \quad \quad \quad \quad -  \sum_{i=1}^{m-1}  ({|a|+\delta \over |a|})^{(\zeta_i - s_i -1)n} ({\alpha \over |a|})^{(1+s_i)n} (2^{R'-1} - 1/2) \Delta_0 \bigg) \bigg\} \nonumber \\
&& \leq \min_{0 \leq m \leq s_{k-1}} \bigg\{ P \bigg\{ |\bar{d}|   \geq \bigg( 2^{R'-1}({\alpha \over |a|})^n \bigg( ({|a|+\delta \over |a|})^{(\zeta_m- s_m - 1)n} ({\alpha \over |a|})^{s_mn} - 2^{-R'} ({\alpha \over |a|})^{-n} \bigg) \Delta_0 \nonumber \\
&& \quad \quad \quad \quad \quad \quad \quad \quad -  \sum_{i=1}^{m-1}  (2^{R'-1}({\alpha \over |a|})^n ({|a|+\delta \over |a|})^{(\zeta_i - s_i -1)n} ({\alpha \over |a|})^{s_in} (1 -  2^{-R'}) \Delta_0) \bigg) \bigg\} \bigg\}
\end{eqnarray}
\setcounter{equation}{\value{mytempeqncnt}}
\hrulefill
\vspace*{4pt}
\end{figure*}
\addtocounter{equation}{1}

As a result (\ref{ohio0}) writes as (\ref{addedLater2}), in this case.
Since $2^{R'-1}({\alpha \over |a|})^n > 1$, the effect of the additional $1$ in the exponent for $({\alpha \over |a|})^{s_m+1}$ can be excluded, unlike the case with $P^g_{e|e}$ above in (\ref{analysisSumBound}).

As a result, the stopping time is dominated by
\begin{eqnarray}\label{keyBoundSum3}
P(\tau_1 \geq kn) 
&\leq& \Xi(\Delta_0) (eP_e^{\kappa})^{k-2},
\end{eqnarray}
for $\Xi(\Delta) = M_r(\Delta) + 1$ which goes to 1 as $\Delta \to \infty$.
\end{itemize}

%
%
%
%
%

This completes the proof of the Lemma.

\qed

\subsection{Proof of Theorem \ref{RecurrenceChannelNoAssumption}:}\label{SufficiencyRecurrence}

Once we have the Markov chain by Lemma \ref{MarkovChain}, and the bound on the distribution of the sequence of stopping times defined in (\ref{stopTimes}) by Lemma \ref{keyLemma}, we will invoke Theorem \ref{thm5} or Theorem \ref{thm26} with Lyapunov functions $V(x,\Delta)=\log_2(\Delta^2)$, $f(x,\Delta)$ taken as a constant and $C$ a compact set.

As mentioned in Remark \ref{RandomCodeBound0}, for a DMC with block length $n$ Shannon's random coding method satisfies:
\begin{eqnarray*}
&& P_e(n): = \max_{c \in \{1,2,\dots, M(n) \}} P(c' \neq c | c \quad \mbox{is transmitted}) \nonumber \\
&& \quad \quad \quad \quad \quad \leq e^{-n E (R) + o(n)},
\end{eqnarray*}
with $c'$ being the decoder output. Here ${o(n) \over n}\to 0$ as $n \to \infty$ and $E(R) > 0$ for $0 < R < C$. Thus, by Lemma \ref{keyLemma}, we observe that,
\begin{eqnarray}\label{uniformBoundStoppingTime}
&& E[\tau_1 | x_0,\Delta_0]  = \sum_{k=1}^{\infty} P(\tau_1 \geq k) \leq K'_{\Delta_0}(n) < \infty,
\end{eqnarray}
for some finite number $K'_{\Delta_0}(n)$. The finiteness of this expression follows from the observation that for $k-2 > {1-\kappa \over \kappa}$, the exponent in $e^{-n (\kappa - {1 -\kappa \over k-2})(E(R) - {o(n) \over n})}$ becomes negative. Furthermore, $K'_{\Delta_0}(n)$ is monotone decreasing in $\Delta_0$ since $M_r(\Delta)$ is decreasing in $\Delta$.


We now apply the random-time drift result in Theorem \ref{thm5} and Corollary \ref{corol} below. First, observe that, the probability that $\tau_{z+1} \neq \tau_z+ n$, is upper bounded by the probability below:
\begin{eqnarray}\label{useful2}
 && P^e_{g|g} \nonumber\\
  && + (1- P^e_{g|g}-P^e_{{\cal Z}|g}) 2 P \bigg( \bar{d} \geq (2^{R'}({\alpha \over |a|})^n -1) \Delta_0/2 \bigg) \nonumber\\
  && \quad + 2 P^e_{{\cal Z}|g} P \bigg( \bar{d} > (2^{R'-1}(|a|+\delta)^n) \Delta_0 - |a^nx_0| \bigg)\nonumber \\
 && \leq P^e_{g|g} \nonumber\\
  && + (1- P^e_{g|g}-P^e_{{\cal Z}|g}) 2 P \bigg( \bar{d} \geq (2^{R'}({\alpha \over |a|})^n -1) \Delta_0/2 \bigg) \nonumber\\
  && \quad + P^e_{{\cal Z}|g} 2 P \bigg( {\bar d} > 2^{R'-1} ((|a|+\delta)^n - |a|^n) \Delta_0 \bigg) \nonumber \\
&& =:\Upsilon(\Delta_{\tau_0})\label{boundTime1}
 \end{eqnarray}
Observe that, provided that $R'(n) > n\log_2(|a|/\alpha)$, \[\lim_{\Delta_0 \to \infty} \Upsilon(\Delta_{\tau_0}) = P^e_{g|g}.\]

\begin{figure*}[!t]
\normalsize
\setcounter{mytempeqncnt}{\value{equation}}
\setcounter{equation}{44}
\begin{eqnarray}\label{geoDrift1}
&& E[\log(\Delta_{\tau_{z+1}}^2)| x_{\tau_{z}},\Delta_{\tau_{z}}] \nonumber \\
&& =E[\log(\Delta_{\tau_{z+1}}^2) (1_{\{\mbox{Type I-A error at $\tau_z$}\}} + 1_{\{\mbox{Type I-B error at $\tau_z$}\}}+ 1_{\{\mbox{no  error at $\tau_z$}\}})| x_{\tau_{z}},\Delta_{\tau_{z}}]\nonumber \\
&& \leq (1 - P^e_{{\cal Z}|g} - P^e_{g|g}) \bigg( n  \log_2(\alpha) + n E[ \log_2((|a+\delta)^{2(\tau_{z+1}-1)}) 1_{\{\tau_{z+1}> \tau_z +n\}}) | \mbox{no error}] \bigg) \nonumber \\
&& \quad \quad + P^e_{{\cal Z}|g} \bigg(n \log_2(|a|+\delta) + n E[ \log_2((|a+\delta)^{2(\tau_{z+1}-1)}) 1_{\{\tau_{z+1}> \tau_z +n\}}) | \mbox{Type I-B error}] \bigg) \nonumber \\
&& \quad \quad + P^e_{g|g}\bigg( n \log_2(\alpha) + n E[\log_2((|a+\delta)^{2(\tau_{z+1}-1)}) 1_{\{\tau_{z+1}> \tau_z +n\}}) | \mbox{Type I-A error}] \nonumber \\
&& \quad \quad+ \log_2(\Delta_{\tau_z}^2) \nonumber \\
&& = \log(\Delta^2_{\tau_{z}}) + n\bigg( (1 - P^e_{{\cal Z}|g}) \log_2(\alpha) + P^e_{{\cal Z}|g} \log_2(|a|+\delta) \bigg) \nonumber \\
&& \quad \quad + n E\bigg[ \log_2\bigg((|a+\delta)^{2(\tau_{z+1}-1)}\bigg) 1_{\{\tau_{z+1}> \tau_z +n\}})\bigg] \nonumber \\
&& \leq \log(\Delta^2_{\tau_{z}}) + n\bigg( (1 - P^e_{{\cal Z}|g}) \log_2(\alpha) + P^e_{{\cal Z}|g} \log_2(|a|+\delta) \bigg) \nonumber \\
&& \quad \quad + \bigg(P(\tau_{z+1} > \tau_z+n)\bigg)^{\chi \over 1+\chi} n \bigg( \sum_{k=2}^{\infty} P(\tau_{z+1}=\tau_z+kn) ((k-1) \log_2(|a|+\delta))^{1+\chi} \bigg)^{1 \over 1+\chi} \label{intermediate} \\
&& \leq \log(\Delta^2_{\tau_{z}}) + n\bigg( (1 - P^e_{{\cal Z}|g}) \log_2(\alpha) + P^e_{{\cal Z}|g} \log_2(|a|+\delta) \bigg) \nonumber \\
&& \quad \quad + (\Upsilon(\Delta_{\tau_z}))^{\chi \over 1+\chi} n \bigg( \sum_{k=2}^{\infty} P(\tau_{z+1}=\tau_z+kn) ((k-1) \log_2(|a|+\delta))^{1+\chi} \bigg)^{1 \over 1+\chi} \label{intermediate2}
\end{eqnarray}
\setcounter{equation}{\value{mytempeqncnt}}
\hrulefill
\vspace*{4pt}
\end{figure*}
\addtocounter{equation}{2}

We now pick the Lyapunov function $V(x,\Delta)= \log_2(\Delta^2)$ and $f(x,\Delta)$ a constant to obtain (\ref{intermediate2}), where $\chi >0$ is an arbitrarily small positive number. In (\ref{intermediate}) we use the fact that zooming out for all time stages after $\tau_z +n$ provides a worst case sequence and that by H\"older's inequality for a random variable $X$ and an event $\mathbb{A}$ the following holds:
\begin{eqnarray}
&& E[X 1_{\mathbb{A}}] \nonumber \\
&& \leq (E[|X|^{1+\chi}])^{{1 \over 1+\chi}} (E[1^{{1+\chi \over \chi}}_{\mathbb{A}}])^{\chi \over 1+\chi} \nonumber \\
&& = (E[|X|^{1+\chi}])^{{1 \over 1+\chi}} (P(\mathbb{A}))^{\chi \over 1+\chi}.
\end{eqnarray}
Now, the last term in (\ref{geoDrift1}) will converge to zero with $n$ large enough and $\Delta_{\tau_z} \to \infty$ for some $\chi > 0$, since by Lemma \ref{keyLemma} $P(\tau_{z+1}=\tau_z+kn)$ is bounded by a geometric measure and the expectation of $((\tau_{z+1}-\tau_z-1) \log_2(|a|+\delta))^{1+\chi}$ is finite and monotone decreasing in $\Delta_0$. The second term in (\ref{intermediate2}) is negative with $P^e_{{\cal Z}|g}$ sufficiently small.

These imply that, for some sufficiently large $F$, the equation
\begin{eqnarray}\label{geoDrift2}
E[\log(\Delta_{\tau_{z+1}}^2)| \Delta_{\tau_{z}}, h_{\tau_{z}}] \leq \log(\Delta_{\tau_{z}}^2) - b_0 + b_1 1_{\{|\Delta_{\tau_z}| \leq F\}}
\end{eqnarray}
holds for some positive $b_0$ and finite $b_1$. Here, $b_1$ is finite since $K'(n)$ is finite. With the uniform boundedness of (\ref{uniformBoundStoppingTime}) over the sequence of stopping times, this implies by Theorem \ref{thm26} that $\{(x, \Delta) : |\Delta_{\tau_z}| \leq F, |{x \over 2^{R'-1}\Delta}| \leq 1\}$ is a recurrent set. \qed


\subsection{Proof of Theorem \ref{Cyclostationary}} \label{ProofInv}
The process $(x_{tn},\Delta_{tn})$ is a Markov chain, as was observed in Lemma \ref{MarkovChain}.
In this section, we establish irreducibility of this chain and the existence of a small set (see Section \ref{RandomDriftTheoremSection}) to be able to invoke Theorem \ref{thm5}, in view of (\ref{geoDrift2}). The following generalizes the analysis in \cite{YukTAC2010} and \cite{YukMeynTAC2010}.

Let the values taken by \[\log_2({\bar Q}(\Delta_{tn},c'_{(t+1)n-1}))/s\] be $\{-\tilde{A},0,\tilde{B}\}$. Here $\tilde{A}, B$ are relatively prime. 
Let $\mathbb{L}_{z_0,\tilde{A},\tilde{B}}$ be defined as $$\{n \in \mathbb{N}, n \geq \log_2({L'})/s: \exists N_A, N_B, n= -N_A \tilde{A} + N_B \tilde{B} + z_0 \},$$
where $z_0 = \log_2(\Delta_{0})/s$ is the initial condition of the parameter for the quantizer. We note that since $\tilde{A}, \tilde{B}$ are relatively prime, by B\'ezout's Lemma (see \cite{NumberTheory}) the communication class will include the bin sizes whose logarithms are integer multiples of a constant except those leading to $\Delta < L'$: Since we have $\Delta_{(t+1)n}={\bar Q}(\Delta_{tn},c'_{(t+1)n-1}) \Delta_{tn}$, it follows that
 $$\log_2(\Delta_{(t+1)n})/s = \log_2({\bar Q}(\Delta_{tn},c'_{(t+1)n-1}))/s + \log_2(\Delta_{tn})/s,$$
 is also an integer. Furthermore, since the source process $\{x_{tn}\}$ is ``Lebesgue-irreducible'' (the system noise admits a probability density function that is positive everywhere), and there is a uniform lower bound $L'$ on bin-sizes, the error process takes values in any of the admissible quantizer bins with non-zero probability. Consider two integers $k,l \geq {\log_2(L') \over s}$. For all $l,k \in \mathbb{L}_{z_0,\tilde{A},\tilde{B}}$, there exist $N_A, N_B \in \mathbb{N}$ such that $l-k = - N_A \tilde{A} + N_B \tilde{B}.$
We can show that the probability of $N_A$ occurrences of perfect zoom, and $N_B$ occurrences of under-zoom phases is bounded away from zero. This set of occurrences includes the event that in the first $N_A$ time stages perfect-zoom occurs and later, successively, $N_B$ times under-zoom phase occurs. Considering worst possible control realizations and errors, the probability of this event is lower bounded by
\begin{eqnarray}\label{MoveBin2Bin2}
&& \bigg(\Prob\bigg({\tilde d} \in [-2^{R'(n)-1}L' - |a|^nL', 2^{R'(n)-1}L' -  |a|^nL']\bigg) \nonumber \\
&& \quad \quad \quad \quad \times (P^e({\cal Z}|i)) \bigg)^{N_B} \nonumber \\
&& \quad \times \bigg(\Prob\bigg(\tilde{d} \in [-(\alpha^n 2^{R'} - a^n)L'/2, (\alpha^n 2^{R'} - a^n)L'/2]) \nonumber \\
&& \quad \quad \quad \quad \quad \times (1-P_e) \bigg)^{N_A} > 0,
\end{eqnarray}
where $\tilde{d} = \sum_{i=0}^{n-1} a^{n-i-1} w_i$ is a Gaussian random variable. The above follows from considering the sequence of zoom-ins and zoom-outs and the behavior of $a^n(x_{tn}- \hat{x}_{tn}) + \tilde{d}$. In the above discussion, $P^e({\cal Z}|i)$ is the conditional error on the zoom symbol given the transmission of granular bin $i$, with the lowest error probability (If the lowest such an error probability is zero, an alternative sequence of events can be provided through events concerning the noise variables leading to zooming). Thus, for any two such integers $k,l$ and for some $r > 0$, $\Prob(\log_2(\Delta_{(t+r)n}) = ls \mid \log_2(\Delta_{tn})=ks ) > 0.$

We can now construct a small set and make the connection with Theorems \ref{RecurrenceChannelNoAssumption} and ~\ref{thm5}. Define
  \[C_x \times C'_{\Delta} = \{(x, \Delta): L' \leq \Delta \leq F, |h| \leq 1, {\log_2(\Delta) \over s} \in \mathbb{Z} \} \]
We will show that the recurrent set $C_x \times C'_{\Delta}$ is small.

Towards this end, we first establish irreducibility. For some distribution ${\cal K} $ on positive integers, $E \subset \mathbb{R}$ and $\Delta$ an admissible bin size,
\begin{eqnarray}\label{smalldiscussion}
&& \sum_{n \in \mathbb{N}_+}{\cal K}(n) \Prob\bigg((x_n,\Delta_n) \in (E \times \{\Delta\}) \,\Big|\,   x_0,\Delta_0 \bigg) \nonumber \\
&& \quad \quad \geq K_{\Delta_0,\Delta} \psi(E,\Delta) \nonumber
\end{eqnarray}

%

Here $K_{\Delta_0,\Delta}$, denoting a lower bound on the probability of visiting $\Delta$ from $\Delta_0$ in some finite time, is non-zero by (\ref{MoveBin2Bin2}) and $\psi$ is a positive map as the following argument shows. Let $t>0$ be a time stage for which $\Delta_{tn}=\Delta$ and thus, with $|h_{(t-1)n}| \leq 1$: $|ax_{(t-1)n}+bu_{(t-1)n}| \leq |a|^n\Delta_{(t-1)n}/2 = (|a|/\alpha)^n { \Delta \over 2}$. Thus, it follows that, for $A_1, B_1 \in \mathbb{R}$, $A_1 < B_1$,
\begin{eqnarray}
&& \Prob\bigg(x_{tn} \in [A_1, B_1] \,\Big|\,   |a^nx_{(t-1)n}+bu_{(t-1)n}| \nonumber \\
&& \quad \quad \quad \quad \quad \leq |a|\Delta_{(t-1)n}/2 \bigg) \nonumber \\
&& = \Prob\bigg(\tilde{d} _{t-1} \in [A_1 - (a^nx_{(t-1)n}+bu_{(t-1)n})\nonumber \\
&& \quad \quad \quad \quad \quad \quad \quad , B_1 - (a^nx_{(t-1)n}+bu_{(t-1)n})]\nonumber\\
&& \quad \quad \quad \,\Big|\,   |a^nx_{(t-1)n}+bu_{(t-1)n} | \leq |a|^n\Delta_{(t-1)n}/2 \bigg) \nonumber \\
&& \geq \min_{|z| \leq { \Delta \over 2} (|a|/\alpha)^n} \bigg( \Prob(\tilde{d} _{t-1} \in [A_1 - z, B_1 - z] \bigg)
\end{eqnarray}

Thus, in view of (\ref{smalldiscussion}), $\psi$ satisfies for $\alpha<\beta, \alpha,\beta \in \mathbb{R},$
\begin{eqnarray}
&& \psi([A_1,B_1],\Delta) \nonumber \\
&& \geq \min_{|z| \leq { \Delta \over 2} (|a|/\alpha)^n} \bigg( \Prob(\tilde{d}_{t-1} \in [A_1 - z, B_1 - z] \bigg) > 0 \nonumber
\end{eqnarray}

The chain satisfies the recurrence property that
\[\Prob_{(x,\Delta)}(\tau_{C_x \times C'_{\Delta}} < \infty  ) =1,\] for any admissible $(x, \Delta)$. This follows from the construction of
\[\Theta_k(\Delta,x) := \Prob(\stp_{1} \geq kn \mid x,\Delta), \]
where
\[\stp_{1} = \inf(kn > 0: |x_k| \leq 2^{R'-1}\Delta_k, x_0=x, \Delta_0=\Delta )\]
and observing that $\Theta_k(\Delta,x)$ is majorized by a geometric measure with similar steps as in \Section{ProofStocStabSystemTimeDistribution}. Once a state which is perfectly zoomed, that is which satisfies $|x_t| \leq 2^{R'-1}\Delta_t$, is visited, the stopping time analysis can be used to verify that from any initial condition the recurrent set is visited in finite time with probability 1.

 We will now establish that the set $C_x \times C'_{\Delta}$ is small. By Theorem 5.5.7 of \cite{MCSS}, under aperiodicity and irreducibility, every petite set is small. To establish the petite set property, we will follow an approach taken by Tweedie \cite{Tweedie2001} which considers the following test, which only depends on the one-stage transition kernel of a Markov chain: If a set $S$ is such that, the following {\em uniform countable additivity} condition
\begin{eqnarray}
\lim_{k \to \infty} \sup_{x \in S} P(x,B_k) = 0, \nonumber
\end{eqnarray}
 is satisfied for every sequence $B_k \downarrow \emptyset$, and if the Markov chain is irreducible, then $S$ is petite (see Lemma 4 of Tweedie \cite{Tweedie2001} and Proposition 5.5.5(iii) of Meyn-Tweedie \cite{MCSS}).


\begin{figure*}[!t]
\normalsize
\setcounter{mytempeqncnt}{\value{equation}}
\setcounter{equation}{50}
\begin{eqnarray}\label{uniformcountableadditive}
&& \lim_{k \to \infty} \sup_{(x, \Delta) \in C_x \times C'_{\Delta}} \Prob((x_{(t+1)n},\Delta_{(t+1)n}) \in (B_k \times \Delta') | x_{tn}=x,\Delta_{tn}=\Delta) \nonumber \\
&& =  \lim_{k \to \infty} \sup_{(x, \Delta) \in C_x \times C'_{\Delta}} \Prob\bigg((\tilde{d},\Delta_{(t+1)n}) \in \bigg((B_k - (a^nx_{tn} + bu_{(t+1)n-1})) \times \Delta'\bigg) \bigg| x_{tn}=x,\Delta_{tn}=\Delta\bigg) = 0.
 \end{eqnarray}
\setcounter{equation}{\value{mytempeqncnt}}
\hrulefill
\vspace*{4pt}
\end{figure*}
\addtocounter{equation}{1}

Now, the set $C_x \times C'_{\Delta}$ satisfies the uniform countable additivity condition since for any given bin size $\Delta'$ in the countable space constructed above, (\ref{uniformcountableadditive}) holds. This follows from the fact that the Gaussian random variable $\tilde{d}$ satisfies \[\lim_{k \to \infty} \sup_{\tilde{d} \in C_0} \Prob(\tilde{d} \in A_k) = 0,\] uniformly over a compact set $C_0$, for any sequence $A_k \downarrow \emptyset$, since a Gaussian measure admits a uniformly bounded density function. Hence, $C_x \times C'_{\Delta}$ is petite.

Finally, aperiodicity of the sampled chain follows from the fact that the smallest admissible state for the quantizer, $L'$ can be visited in subsequent time sampled time stages, since \[P(\tilde{d} \in [-2^{R'-1}L'/|a|^n- L', -2^{R'-1}L'/|a|^n + L']) >0.\]

Thus, the sampled chain is positive Harris recurrent.

%
%
%
%
%
%
%

 \qed

\subsection{Proof of Theorem \ref{ErgodicityImpliedSampled}} \label{ProofErgodicityImpliedSampled}
By Kolmogorov's Extension Theorem, it suffices to check that the property holds for finite dimensional cylinder sets, since these sets generate the $\sigma-$algebra on which the stochastic process measure is defined. Suppose first that the sampled Markov chain is stationary. Consider two elements:
\begin{eqnarray}
&&P(x_{t+1+n} \in A_1,x_{t+2+n} \in A_2) \nonumber \\
&& = \int_{x_{\lfloor {t+1+n \over n} \rfloor n}}P(dx_{{\lfloor {t+1+n \over n} \rfloor n}}, x_{t+1+n} \in A_1,x_{t+2+n} \in A_2)\nonumber \\
&&= \int_{x_{\lfloor {t+1+n \over n} \rfloor n}} P(x_{t+1+n} \in A_1,x_{t+2+n} \in A_2| x_{\lfloor {t+1+n \over n} \rfloor n}) \nonumber \\
&& \quad \quad \quad  \quad \quad \quad \times P(dx_{\lfloor {t+1+n \over n} \rfloor n} )\nonumber \\
&&= \int_{x_{\lfloor {t+1 \over n} \rfloor n}} P(x_{t+1} \in A_1 ,x_{t+2} \in A_2| x_{\lfloor {t+1 \over n} \rfloor n}) P(dx_{\lfloor {t+1 \over n} \rfloor n} ) \nonumber
\end{eqnarray}
The above follows from the fact that, the marginals $P(dx_{\lfloor {t+1 \over n} \rfloor n})$ and $P(dx_{\lfloor {t+1+n \over n} \rfloor n})$ are equal since the sampled Markov chain is positive Harris recurrent and assumed to be stationary, and the dynamics for inter-block times are time homogeneous Markov. The above is applicable for any finite dimensional set, thus for any element in the sigma field generated by the finite dimensional sets, on which the stochastic process is defined. Now, let for some event $A$, $T^{-n}A = A$, where $T$ denotes the shift operation (see Section \ref{ergodicTheory}). Then
\[P(A) = \lim_{k\to\infty} P(A \cap T^{-kn}A) = \lim_{k\to\infty} P(A)P(T^{-kn}A|A)\]
Note that a positive Harris recurrent Markov chain admits a unique invariant distribution and for every $x_0 \in \mathbb{R}$, \[\lim_{k \to \infty}P(x_{kn} \in A | x_0) = \pi(A),\] where $\pi(\cdot)$ is the unique invariant probability measure.
Since such a Markov chain forgets its initial condition, it follows that for $A=T^{-n}A$:
\[P(A \cap T^{-kn}A) = P(A \cap A) = P(A)P(A),\]
thus, $P(A) \in \{0,1\}$, and the process is $n-$ergodic. \qed


\subsection{Proof of Theorem \ref{FiniteMomentChannel}}\label{ProofFiniteMoment}
We begin with the following result, which is a consequence of Theorem \ref{thm5}:
\begin{lem} \label{boundState}
Under the conditions of Theorem \ref{Cyclostationary}, we have that, if for some $\gamma > 0$, $b < \infty$, the following holds
\begin{eqnarray}
&& \gamma E[\sum_{k=0}^{(\tau_1/n)-1} \Delta_{kn}^2 | x_0, \Delta_0] \nonumber \\
&& \leq \Delta_0^2 - E[\Delta_{\tau_1}^2|x_0, \Delta_0] + b 1_{\{(\Delta_0,h_0) \in (C'_x \times C_h)\}},\nonumber
\end{eqnarray}
then, $\lim_{k \to \infty} E[\Delta_{kn}^2] < \infty.$
\qed \end{lem}


\begin{figure*}[!t]
\normalsize
\setcounter{mytempeqncnt}{\value{equation}}
\setcounter{equation}{51}
\begin{eqnarray}\label{Number1}
&& \Expect[\sum_{t=0}^{(\tau_1/n)-1} \Delta_{tn}^2 \mid x_0, \Delta_0] \nonumber \\
&&\leq \Delta_0^2 P^e_{g|g} \bigg(\sum_{l=2}^{\infty} P(\tau_1=ln | \mbox{Type I-A error}) \sum_{k=1}^{(l-1)} (|a|+\delta)^{2(k-1)n}\alpha^{2n} \bigg) \nonumber \\
&& \quad \quad + \Delta_0^2 P^e_{{\cal Z}|g} \bigg(\sum_{l=2}^{\infty} P(\tau_1=ln | \mbox{Type I-B error}) \sum_{k=1}^{l-1} (|a|+\delta)^{2kn} \bigg) \nonumber \\
&& \quad \quad + \Delta_0^2 (1 - P^e_{g|g} - P^e_{{\cal Z}|g}) \bigg(\sum_{l=2}^{\infty} P(\tau_1=ln | \mbox{no error at time $0$}) \sum_{k=1}^{l-1} (|a|+\delta)^{2(k-1)n}\alpha^{2n}\bigg) \nonumber \\
&& \quad \quad \quad \quad + \Delta_0^2 P(\tau_1 = n) \nonumber \\
&&\leq \Delta_0^2 P^e_{g|g} \bigg(\sum_{l=2}^{\infty} P(\tau_1=ln | \mbox{Type I-A error}) {(|a|+\delta)^{2(l-1)n} \over (|a|+\delta)^{2n} - 1} \alpha^{2n} \bigg) \nonumber \\
&& \quad \quad + \Delta_0^2 P^e_{{\cal Z}|g} \bigg(\sum_{l=2}^{\infty} P(\tau_1=ln | \mbox{Type I-B error}) (|a|+\delta)^{2n} {(|a|+\delta)^{2(l)n} \over (|a|+\delta)^{2n} - 1}  \bigg) \nonumber \\
&& \quad \quad + \Delta_0^2 (1 - P^e_{g|g} - P^e_{{\cal Z}|g}) \bigg(\sum_{l=2}^{\infty} P(\tau_1=ln | \mbox{no error at time $0$}) {(|a|+\delta)^{2(l-1)n} \over (|a|+\delta)^{2n} - 1} \alpha^{2n}\bigg) \nonumber \\
&& \quad \quad \quad \quad + \Delta_0^2 P(\tau_1 = n) \nonumber \\
&& \leq \Delta_0^2 P^e_{g|g} {(|a|+\delta)^{(2/\kappa)n} \over (|a|+\delta)^{2n} - 1} \Xi(\Delta_0) \bigg( \sum_{l=2}^{\infty} (e^{(l-2)}) P_e^{(\kappa)(l-1-{1 \over \kappa})} (|a|+\delta)^{2n(l-1 - {1 \over \kappa})} \alpha^{2n} \bigg) \nonumber \\
&& \quad \quad + \Delta_0^2 P^e_{{\cal Z}|g} {(|a|+\delta)^{2n} \over 1 - (|a|+\delta)^{-2n}} \Xi(\Delta_0) \bigg(\sum_{l=2}^{\infty}  (eP_e^{\kappa})^{l-2} (|a|+\delta)^{2(l-2)n} \bigg) \nonumber \\
&& \quad \quad + \Delta_0^2 (1 - P^e_{g|g} - P^e_{{\cal Z}|g}) (|a|+\delta)^{2n} \Xi(\Delta_0) \bigg(\sum_{l=2}^{\infty} (eP_e^{\kappa})^{l-2} {(|a|+\delta)^{2(l-2)n} \over (|a|+\delta)^{2n}-1} \alpha^{2n} \bigg) \nonumber \\
&& \quad \quad \quad \quad + \Delta_0^2 P(\tau_1 = n) \nonumber\\
&& \leq \zeta_1 \Delta_0^2
\end{eqnarray}
\setcounter{equation}{\value{mytempeqncnt}}
\hrulefill
\vspace*{4pt}
\end{figure*}
\addtocounter{equation}{1}

Now, under the hypotheses of Theorem \ref{Cyclostationary}, and observing that Type I-B and I-A errors are worse than the no error-case at time $0$ for the stopping time tail distributions, we obtain (\ref{Number1}) for some finite $\zeta_1$. In (\ref{Number1}), we use the property that $H(\kappa) \leq 1$ and (\ref{keyBoundSum1})-(\ref{keyBoundSum3}).

We now establish that
\[\lim_{\Delta_0 \to \infty} {\Expect[ \Delta_{\tau_1}^2 \mid x_0, \Delta_0] \over \Delta_{0}^2} < 1.\]
This is a crucial step in applying Theorem \ref{thm5}.

Following similar steps as in (\ref{Number1}), the following upper bound on $(\Expect[\Delta_{\tau_1}^2 \mid x_0, \Delta_0] / \Delta_0^2)$ is obtained:
\begin{eqnarray}\label{BoundFatSquirrel}
&& (1- P^e_{g|g} - P^e_{{\cal Z}|g}) \nonumber \\
&& \quad \times \bigg( \alpha^{2n} + {1 \over \Delta_0^2}\Expect[\Delta_{\tau_1}^2 1_{\{\tau_1 > n\}}| \mbox{no error at time $0$}] \bigg) \nonumber \\
&&+ (P^e_{g|g}) \bigg( \alpha^{2n} (1 + (|a|+\delta)^{2n} + \dots (|a|+\delta)^{2(\lfloor{1 \over \kappa} \rfloor)n} ) \nonumber\\
&& \quad + \sum_{k=\lceil {1 \over \kappa} \rceil+1}^{\infty} e^{k-2}P_e^{(\kappa(k-1 - {1 \over \kappa}))} (|a|+\delta)^{2(k-1 - {1 \over \kappa})n} (\alpha)^{2n}   \nonumber \\
&& \quad \quad \quad \times (|a|+\delta)^{(2/\kappa)n} \Xi(\Delta_0) \bigg)  \nonumber\\
&& + P^e_{{\cal Z}|g} \bigg( (|a|+\delta)^{2n} + \Xi(\Delta_0) {P_e^{\kappa}(|a|+\delta)^{2n}  \over 1 - P_e^{\kappa} (|a|+\delta)^{2n} } \bigg)
\end{eqnarray}
%

Note now that \[\lim_{\Delta_{0} \to 0}P(\tau_1 > n | \mbox{no error at time $0$}, x_0, \Delta_0) = 0,\] uniformly in $x_0$ with $|h_0| \leq 1$ and given the rate condition $R'(n) > n \log_2(|a|/\alpha)$ by (\ref{boundTime1}). Therefore, the first term in (\ref{BoundFatSquirrel}) to $0$ in the limit of large $\Delta_0$, since
$\lim_{n \to \infty} \kappa{1 \over n}\log(P_e) + 2\log_2(|a|+\delta) < 0$ and we have the following upper bound
\[\bigg((|a|+\delta)^{2n} \sum_{k=2}^{\infty} (eP_e^{\kappa})^{k-2} (|a|+\delta)^{2n(k-2)} (\alpha)^{2n} \bigg) < \infty,\]
for sufficiently large $n$.


For the second term in (\ref{BoundFatSquirrel}), the convergence of the first expression is ensured with $\lim_{n \to \infty} P^e_{g|g} (|a|+\delta)^{(2/\kappa)n} \alpha^{2n} \to 0$ and that $P_e(|a|+\delta)^{(2/\kappa)n} \to 0$ as $n \to \infty$. By combining the second and the third term, the desired result is obtained.

To show that $\lim_{m \to \infty }E[x_{mn}^2] < \infty$, we first show that for some $\kappa >0$,
\begin{eqnarray}\label{keyLemma1}
\kappa \Expect[\sum_{m=0}^{(\tau_1/n)-1} x_{mn}^2 \mid x_0,\Delta_0] \leq \Delta_0^2 2^{2(R'-1)}.
\end{eqnarray}
Now,
\begin{eqnarray}\label{parallel}
&& \Expect[\sum_{m=0}^{(\tau_1/n)-1} x_{mn}^2 \mid x_0,\Delta_0] \nonumber \\
&& = \Expect \bigg[\sum_{t=0}^{(\tau_1/n)-1} a^{2tn} \bigg( (x_0 + \sum_{i=0}^{tn-1} a^{-i-1} d_i) + (\sum_{i=0}^{tn-1} a^{-i-1}u_i) \bigg)^2 \nonumber \\
&& \quad \quad \quad \quad \quad \quad \bigg| x_0,\Delta_0\bigg] \nonumber \\
&& \leq 2 \Expect[\sum_{t=0}^{(\tau_1/n)-1} a^{2tn}(x_0 + \sum_{i=0}^{tn-1} a^{-i-1} d_i)^2  | x_0,\Delta_0] \nonumber \\
&& \quad \quad \quad + 2 \Expect[ \sum_{t=0}^{(\tau_1/n)-1} a^{2tn}  (\sum_{i=0}^{tn-1} a^{-i-1}u_i)^2 | x_0,\Delta_0],
\end{eqnarray}
which follows from the observation that for $X,Y$ random variables, $E[(X+Y)^2] \leq 2E[X^2]+2E[Y^2]$. 

Let us first consider the component: $(a^{t}(x_0 + \sum_{i=0}^{tn-1} a^{-i-1} d_i))^2$.
\begin{eqnarray}
&& \Expect[\sum_{t=0}^{(\tau_1/n)-1} (a^{tn}(x_0 + \sum_{i=0}^{tn-1} a^{-i-1} d_i))^2  | x_0,\Delta_0]  \nonumber \\
&&= \Expect[\sum_{t=0}^{\infty} 1_{\{t < \tau_1/n \}} (a^{tn}(x_0 + \sum_{i=0}^{tn-1} a^{-i-1} d_i))^2 | x_0,\Delta_0] \nonumber \\
&& \leq \sum_{t=0}^{\infty} \bigg( \Expect[(1_{\{t < \tau_1/n\}})^{1 + \chi}| x_0,h_0] \bigg)^{1 \over 1+ \chi} \nonumber \\
&& \quad \quad \times \bigg(\Expect[ (a^{tn}(x_0 + \sum_{i=0}^{tn-1} a^{-i-1} d_i))^{2({1 + \chi \over \chi})} | x_0,\Delta_0] \bigg)^{\chi \over 1+\chi}
 \label{sonDenklem}
\end{eqnarray}
for some $\chi>0$, by H\"older's inequality.

Moreover, for some $B_2<\infty$,
\begin{eqnarray} \label{BoundHolder2}
&&\Expect[ a^{2tn({1 + \chi \over \chi})}(x_0 + \sum_{i=0}^{tn-1} a^{-i-1}d_i)^{2({1 + \chi \over \chi})} | x_0,\Delta_0]  \nonumber \\
&&  = |a|^{2tn({1 + \chi \over \chi})} \Expect[  (x_0 + \sum_{i=0}^{tn-1} a^{-i-1} d_i))^{2{1 + \chi \over \chi}} | x_0,\Delta_0] \nonumber \\
&& \leq |a|^{2tn({1 + \chi \over \chi})} \Expect[  (x_0 + \sum_{i=0}^{\infty} a^{-i-1} d_i)^{2{1 + \chi \over \chi}} | x_0,\Delta_0] \nonumber \\
&& = |a|^{2tn({1 + \chi \over \chi})} (2^{R'-1}\Delta_0)^{2{1 + \chi \over \chi}} \nonumber \\
&& \quad \quad \quad \times \Expect[  ({x_0 + \sum_{i=0}^{\infty} a^{-i-1} d_i \over 2^{R'-1}\Delta_0})^{2{1 + \chi \over \chi}} | x_0,\Delta_0] \nonumber \\
&& = |a|^{2tn({1 + \chi \over \chi})} (2^{R'-1}\Delta_0)^{2{1 + \chi \over \chi}} \nonumber \\
&& \quad \quad \quad \times \Expect[  (h_0 + {\sum_{i=0}^{\infty} a^{-i-1} d_i \over 2^{R'-1}\Delta_0})^{2{1 + \chi \over \chi}} | x_0,\Delta_0] \nonumber \\
&& < B_2 (2^{R'-1}\Delta_0)^{2{1 + \chi \over \chi}} |a|^{2tn({1 + \chi \over \chi})},
  \end{eqnarray}
where the last inequality follows since for every fixed $|h_0| \leq 1$, the random variable $h_0 + (\sum_{i=0}^{\infty} a^{-i-1} d_i) / (2^{R'-1}\Delta_0)$ has a Gaussian distribution with finite moments, uniform on $\Delta_0 \geq L'$. 

Thus,
\begin{eqnarray}\label{BoundHolder22}
 && \Expect[\sum_{t=0}^{(\tau_1/n)-1} a^{2tn}(x_0 + \sum_{i=0}^{tn-1} a^{-i-1} d_i)^2  | x_0,\Delta_0]  \nonumber \\
 &\leq& \sum_{t=0}^{\infty} \bigg(\Expect[(1_{\{t < \tau_1/n\}})^{1 + \chi}| x_0,\Delta_0] \bigg)^{1 \over 1+ \chi} \nonumber \\
&& \quad \quad \quad \times \bigg(B_2 (2^{R'(n)-1}\Delta_0)^{2{1 + \chi \over \chi}} |a|^{2tn({1 + \chi \over \chi})} \bigg)^{\chi \over 1+\chi} \nonumber \\
&=& \sum_{t=0}^{\infty} \bigg(\Xi(\Delta_0) (eP_e^{(\kappa-{1 - \kappa \over t-1})})^{t-1}\bigg)^{1 \over 1+ \chi} \bigg(B_2^{\chi \over 1+\chi} \Delta_0^2 |a|^{2tn} \bigg) \nonumber\\
&=& \sum_{t=0}^{\infty} \bigg(\Xi(\Delta_0) (eP_e^{(\kappa-{1 - \kappa \over t-1})})^{t-1} |a|^{2tn(1 + \chi)} \bigg)^{1 \over 1+ \chi} \nonumber\\
&& \quad \quad \quad \quad \quad \quad \times \bigg(B_2^{\chi \over 1+\chi} \Delta_0^2 \bigg) \nonumber\\
&<& \zeta_{B_2} \Delta_0^2,
\end{eqnarray}
for some finite $\zeta_{B_2}$ (for a fixed finite $n$). In the discussion above we use the fact that we can pick $\chi >0$ such that $(P_e)^{\kappa}|a|^{2n(1 + \chi)} < 1.$ Such a $\chi$ exists, by the hypothesis that $\lim_{n \to \infty} P_e^{\kappa}(|a| + \delta)^{2n} = 0$.

We now consider the second term in (\ref{parallel}). Since $u_i$ is the quantizer output which is bounded in magnitude in proportion with $\Delta_i$, the second term writes as:
\begin{eqnarray}\label{BoundHolder222}
&& \Expect[ \sum_{t=0}^{(\tau_1/n)-1} a^{2tn}  (\sum_{i=0}^{tn-1} a^{(-i-1)}u_i)^2 | x_0,\Delta_0] \nonumber \\
&&\leq \Expect[ \sum_{t=0}^{(\tau_1/n)-1} a^{2tn}  (\sum_{i=0}^{t-1} a^{-in} 2^{(R'-1)}\Delta_{in} )^2 | x_0,\Delta_0]\nonumber \\
&&
\leq  \Expect[ \sum_{t=0}^{(\tau_1/n)-1} a^{2tn}  (\sum_{i=0}^{t-1} a^{-in} 2^{(R'-1)} (|a|+\delta)^{in} \Delta_0)^2 \nonumber \\
&& \quad \quad \quad \quad \quad \quad | x_0,\Delta_0]\nonumber \\
&&
\leq \Expect[ \sum_{t=0}^{(\tau_1/n)-1} a^{2tn} \bigg( 2^{(R'-1)} ({|a|+\delta \over |a|})^{tn}  \Delta_0 \bigg)^2 | x_0,\Delta_0]  \nonumber \\
&& \quad \quad \quad \times \bigg({1 \over (1 - ({|a|+\delta \over |a|})^{n})}\bigg)^2 \nonumber \\
&& \leq \Delta^2_0 \tilde{\zeta}_{B}
\end{eqnarray}
for some finite $\tilde{\zeta}_{B}$, by the bound on the stopping time and arguments presented earlier.

Now, with (\ref{BoundFatSquirrel}), (\ref{parallel}), (\ref{BoundHolder22}-\ref{BoundHolder222}), we can apply Theorem \ref{thm5}: With some $0 < \epsilon$ (whose existence is justified by (\ref{BoundFatSquirrel})),
  \[\delta(x,\Delta) = \epsilon \Delta^2, \quad f(x,\Delta) = {\epsilon \over 2 \zeta_{B_2} + 2 \tilde{\zeta}_{B}} x^2,\] $C$ a compact set and $V(x,\Delta)=\Delta^2$, Theorem \ref{thm5} applies and $\lim_{t \to \infty} E[x_{tn}^2] < \infty$.

Thus, with average rate strictly larger than $\log_2(|a|)$, stability with a finite second moment is achieved. Finally, the limit is independent of the initial distribution since the sampled chain is irreducible, by Theorem \ref{Cyclostationary}.
Now, if the sampled process has a finite second moment, the average second moment for the state process satisfies
\[\lim_{N \to \infty}{1 \over N} E[\sum_{k=0}^{N-1} x_k^2] = {1 \over n} E_{\pi}[\sum_{k=0}^{n-1} x_k^2|x_0,\Delta_0],\] is also finite, where $E_{\pi}$ denotes the expectation under the invariant probability measure for $x_0,\Delta_0$. By the ergodic theorem for Markov chains (see Theorem \ref{thm5}), the above holds almost surely, and as a result
\[\lim_{N \to \infty}{1 \over N} \bigg(\sum_{k=0}^{N-1} x_k^2\bigg) = {1 \over n} E_{\pi}[\sum_{k=0}^{n-1} x_k^2|x_0,\Delta_0] < \infty \quad a.s.\]
\qed

\subsection{Proof of Theorem \ref{FiniteMomentChannelA1}} \label{ProofFiniteMomentChannelA1}

Proof follows from the observation that the number of errors in channel transmission when the state is under-zoomed, $s$, is zero. No errors take place in the phase when the quantizer is being zoomed out.

Following (\ref{BoundFatSquirrel}), the only term which survives is
\begin{eqnarray}\label{BoundFatSquirrel22}
&& \alpha^{2n} + P^e_{g|g} \bigg( \alpha^{2n} \bigg(1 + (|a|+\delta)^{2n} + \dots (|a|+\delta)^{2(\lfloor {1 \over \kappa} \rfloor)n} \bigg) \nonumber
\end{eqnarray}
which is to be less than $1$. We can take $\kappa> 1/2$ for this case. 
Now,
\begin{eqnarray}
&&\lim_{\Delta \to \infty} P(\tau \geq kn | x_0, \Delta_0) \nonumber \\
&& \quad \quad \leq P(\bar{d} > {(|a|+\delta)^{(k-1)n}\alpha^n \over |a|^{kn}} \Delta_0 2^{R'-1}) = 0, \nonumber
\end{eqnarray}
for $k > {1 \over \kappa}$. Hence,
$\lim_{n \to \infty} P^e_{g|g} (|a|+\delta)^{2n} \to 0$ is sufficient, since $2 > {1 \over \kappa}$. The proof is complete once we recognize $\bar{P}_e$ as $P^e_{g|g}$.

 \qed

\subsection{Proof of Theorem \ref{StochasticStabilityofVector}}\label{ProofForVectorCaseAMS}

We provide a sketch of the proof since the analysis follows from the scalar case, except for the construction of an adaptive vector quantizer and the associated stopping time distribution.

Consider the following system
\begin{eqnarray}\label{MultiDimSystem2}
\begin{bmatrix} x^1_{t+1} \\ x^2_{t+1} \\ \vdots \\ x^n_{t+1} \end{bmatrix} = \Lambda \begin{bmatrix} x^1_{t} \\ x^2_{t} \\ \vdots \\ x^n_{t} \end{bmatrix} + \tilde{B} u_t + \tilde{G} d_t,
\end{eqnarray}
where $\Lambda=\mbox{Diag}(\lambda^i)$ is a diagonal matrix, obtained via a similarity transformation: $\Lambda = U^{-1} A U$ and $\tilde{B}=U^{-1}B, \tilde{G}=U^{-1}G$, where $U$ consists of the eigenvectors of the matrix $A$. We can assume that, without any loss, $\tilde{B}$ is invertible since otherwise, by the controllability assumption, we can sample the system with a period of at most $N$ to obtain an invertible control matrix.

The approach now is quantizing the components in the system according to the adaptive quantization rule provided earlier, except for a joint mapping for the overflow region. We modify the scheme in (\ref{QuantizerUpdate2}) as follows: Let for $i=1,2,\dots,n$, $R'_i(n)=\log_2(2^{R_i(n)}-1)=\log_2(K_i(n))$. The vector quantizer quantizes uniformly the marginal variables and we define the overflow region as the quantizer outside the granular region: $\prod_{i=1}^N [-2^{R'_i(n)-1} \Delta^i, 2^{R'_i(n)-1} \Delta^i]$ and for $ \quad i=1,2,\dots,N$
\[Q^{\Delta^i_t}_{K_i}(x)={\cal Z} \quad \mbox{if} \quad x \notin \prod_{k=1}^N [-2^{R'_k(n)-1} \Delta^k, 2^{R'_k(n)-1} \Delta^k]\]
and for $x \in \prod_{i=1}^N [-2^{R'_i(n)-1} \Delta^i, 2^{R'_i(n)-1} \Delta^i]$, the quantizer quantizes the marginal according to (\ref{UniformQuantizerRule}).
Hence, here $\Delta^i$ is the bin size of the quantizer in the direction of the eigenvector $x^i$, with rate $R'_i(n)$. For $1 \leq i \leq N$:
\begin{eqnarray}
\label{QuantizerUpdate4}
&&u_t = - 1_{\{t=(k+1)n-1\}} \tilde{B}^{-1} \Lambda^n \hat{x}_{kn}, \nonumber \\
&&\hat{x}^i_t= Q_{K_i}^{\Delta^i_t}(x^i_t), \nonumber \\
&&\Delta^i_{t+1} = \Delta^i_t \bar{Q}^i(\Delta^i_t,c'_{(t+1)n-1}),
\end{eqnarray}
with $\delta^i > 0$, $\alpha^i < 1$ and $L^i > 0$ such that
\begin{eqnarray}
 \bar{Q}^i(\Delta^i,c') =(|\lambda^i| + \delta)^n \quad && \mbox{if } \quad c' = {\cal Z} \nonumber \\
 \bar{Q}^i(\Delta^i,c') =(\alpha^i)^n \quad && \mbox{if } \quad c'  \neq {\cal Z}, \Delta \geq L^i, \nonumber \\
 \bar{Q}^i(\Delta^i,c') = 1 \quad \quad  && \mbox{if } \quad c'  \neq {\cal Z}, \Delta < L^i, \nonumber
\end{eqnarray}
and $R'_i(n) > n\log_2(|\lambda^i|/\alpha^i)$.

Instead of (\ref{stopTimes}), the sequence of stopping times is defined as follows. With $\tau_0 = 0$, define
\begin{eqnarray}
\tau_{z+1} = \inf \{kn > \tau_z : |h^i_{kn}| \leq  1, i=1,2\dots,N \}, \quad k, z \in \mathbb{Z}_+, \nonumber
\end{eqnarray}
where $h^i_t = { x^i_t \over \Delta^i_t 2^{R'_i-1}}$.
Now, we observe that $N-$dimensional system:
\begin{eqnarray}
&& P(\tau_1 > kn | x_0, \Delta_0) \nonumber \\
&& = P\bigg(\bigcap_{t=1}^k \{\bigcup_{i=1}^N (|h^i_t| > 1)\} \mid x_0, \Delta_0 \bigg)\nonumber \\
&& \leq P\bigg( \bigcup_{i=1}^N (|h^i_{kn}| > 1) | \mbox{zoom until}\quad k) \mid x_0, \Delta_0 \bigg) \label{unionBound00} \\
&& \leq \sum_{i=1}^N P(|h^i_{kn}| > 1 | \mbox{zoom until}\quad k, x_0, \Delta_0 )  \label{unionBound1}
\end{eqnarray}
where we apply chain rule for probability in (\ref{unionBound00}) and the union bound in (\ref{unionBound1}). However, for each of the dimensions, $P(|h^i_{kn}| > 1 | \mbox{zoom until}\quad kn, x_0, \Delta_0)$ is dominated by an exponential measure, and so is the sum. Furthermore, $P(\tau_1 > n | x_0, \Delta_0)$ still converges to $0$ provided the rate condition $R'_i(n) > \log_2(|\lambda^i|/\alpha^i)$ is satisfied for every $i$, since $P(\tau_1 > n | x_0, \Delta_0) \leq \sum_{i=1}^N P(|h^i_n| > 1 | x_0, \Delta_0)$. Therefore, analogous results to (\ref{geoDrift1})-(\ref{geoDrift2}) are applicable. Once one imposes a countability condition for the bin size spaces as in Theorem \ref{Cyclostationary}, the desired ergodicity properties are established. \qed

\section{Concluding Remarks}

The paper considered stochastic stabilization of linear systems driven by unbounded noise over noisy channels and established conditions for asymptotic mean stationarity. The conditions obtained are tight with an achievability and a converse. The paper also obtained conditions for the existence of finite second moments. When there is unbounded noise, the result we obtained for the existence of finite second moments required further conditions on reliability for channels when compared with the bounded noise case considered by Sahai and Mitter. We do not have a converse theorem for the finite second moment discussion; it would be interesting to obtain a complete solution for this setup.

We observed in the development that, three types of errors were critical. These bring up the importance of unequal error coding schemes with feedback. Recent results in the literature \cite{Borade} have focused on fixed length schemes without feedback, and variable length with feedback and further research could be useful for networked control problems.

The value of information channels in optimization and control problems (beyond stabilization) is an important problem in view of applications in networked control systems. Further research from the information theory community for non-asymptotic coding results will provide useful applications and insight for such problems. These can also be useful to tighten the conditions for the existence of finite second moments. Moderate channel lengths \cite{Polyanskiy1}, \cite{Polyanskiy2}, \cite{HariSahai}, \cite{SahaiDelayBlock}, and possible presence of noise in feedback \cite{DraperSahai} are crucial issues needed to be explored better in the analysis and in the applications of random-time state-dependent drift arguments \cite{YukMeynTAC2010}.

Finally, we note that the assumption that the system noise is Gaussian can be relaxed. For the second moment stability, a sufficiently light tail which would provide a geometric bound on the stopping times as in (\ref{keyBoundSum1}) through (\ref{ilkBound}) will be sufficient. For the AMS property, this is not needed. For a noiseless DMC, \cite{YukTAC2010} established that a finite second moment for the system noise is sufficient for the existence of an invariant probability measure. We require, however, that the noise admits a density which is positive everywhere for establishing irreducibility.

\subsection{Variable Length Coding and Agreement over a Channel}

Let us consider a channel where, agreement on a binary event in finite time is possible between the encoder and the decoder. By binary events, we mean for example, synchronization of encoding times and agreement on zooming times. It turns out that if the following assumption holds, then such agreements are possible in finite expected time: The channel is such that there exist input letters $x_1, x_2, x_3, x_4$ where $D(P(\cdot|x_1)||P(\cdot|x_2)) = \infty$ and $D(P(\cdot|x_3)||P(\cdot|x_4)) = \infty$. Here, $x_1$ can be equal to $x_4$ and $x_2$ can be equal to $x_3$. For example, the erasure channel satisfies this property. Note that, the above condition is weaker than having a non-zero zero-error capacity, but stronger than what Burnashev's \cite{Burnashev}, \cite{YamamotoItoh} method requires; since there are more hypotheses to be tested.

In such a setting, one could use variable length encoding schemes. Such a design will allow the encoder and the decoder to have transmission in three phases: Zooming, transmission, and error confirmation. Using random-time, state-dependent stochastic drift, we may find alternative schemes for stochastic stabilization.

\section{Appendix: Stochastic Stability of Dynamical Systems}\label{sectionergodicandMarkovreview}

\subsection{Stationary, Ergodic, and Asymptotically Mean Stationary Processes}\label{ergodicTheory}

In this subsection, we review ergodic theory, in the context of information theory (that is with the transformations being specific to the shift operation). A comprehensive discussion is available in Shields \cite{Shields} and Gray \cite{GrayProbabilit}, \cite{GrayKieffer}.

Let $\mathbb{X}$ be a complete, separable, metric space. Let ${\cal B}(\mathbb{X})$ denote the Borel sigma-field of subsets of $\mathbb{X}$. Let $\Sigma=\mathbb{X}^{\infty}$ denote the sequence space of all one-sided or two-sided infinite sequences drawn from $\mathbb{X}$. Thus, for a two-sided sequence space if $x \in \Sigma$ then $x = \{\dots,x_{-1},x_0,x_1,\dots\}$ with $x_i \in \mathbb{X}$. Let $X_n:\Sigma \to \mathbb{X}$ denote the coordinate function such that $X_n(x)=x_n$. Let $T$ denote the shift operation on $\Sigma$, that is $X_n(Tx)=x_{n+1}$. That is, for a one-sided sequence space $T(x_0,x_1,x_2,\dots)=(x_1,x_2,x_3,\dots)$.

Let ${\cal B}(\Sigma)$ denote the smallest sigma-field containing all cylinder sets of the form $\{x: x_i \in B_i, m \leq i \leq n\}$ where $B_i \in {\cal B}(\mathbb{X})$, for all integers $m,n$. Observe that $\cap_{n \geq 0} T^{-n}{{\cal B}(\Sigma)}$ is the tail $\sigma-$field: $\cap_n \sigma(x_n,x_{n+1},\cdots)$, since $T^{-n}(A) = \{x: T^nx \in A\}$.

Let $\mu$ be a stationary measure on $(\Sigma,{\cal B}(\Sigma))$ in the sense that $\mu(T^{-1}B)=\mu(B)$ for all $B \in {\cal B}(\Sigma)$. The sequence of random variables $\{x_n\}$ defined on the probability space $(\Sigma,{\cal B}(\Sigma),\mu)$ is a stationary process.

Such a process is aperiodic if $\mu(\{x: T^{-n}x=x\}) = 0$ for each integer $n$.

%
%
%
%
%

\begin{defn} Let $P$ be the measure on a process. This random process is ergodic if $A=T^{-1}A$ implies that $P(A) \in \{0,1\}$.
\end{defn}

That is, the events that are unchanged with a shift operation are trivial events.
%
%
%
Mixing is a sufficient condition for ergodicity. Thus, a source is ergodic if $\lim_{n \to \infty} P(A \cap T^{-n}B) = P(A) P(B)$, since the process forgets its initial condition. Thus, when one specializes to Markov sources, we have the following: A positive Harris recurrent Markov chain is ergodic, since such a process is mixing and stationary. We will discuss this further in the next section.
\begin{defn} A random process is $N-$stationary, (cyclo-stationary or periodically stationary with period $N$) if the process measure $P$ satisfies $P(T^{-N}B)=P(B)$ for all $B \in {\cal B}(\Sigma)$, or equivalently for any $n \in \mathbb{N}$ samples $t_1, t_2, \dots, t_n$:
\begin{eqnarray}
&&P(x_{t_1} \in A_1, x_{t_2} \in A_2, \dots,x_{t_n} \in A_n)  \nonumber \\
&& \quad = P(x_{t_1 + N} \in A_1, x_{t_2 + N} \in A_2, \dots,x_{t_n + N} \in A_n) \nonumber
\end{eqnarray}
\end{defn}
\begin{defn} A random process is $N-$ergodic if $A=T^{-N}A$ implies that $P(A) \in \{0,1\}$.
\end{defn}

\begin{defn}\label{CoordinateRecurrent} A set $A \in {\cal B}(\mathbb{X})$ is coordinate-recurrent if for some $m \in \mathbb{Z}_+$
\[\sum_{m=0}^{\infty} 1_{\{X_m(x) \in A \}} = \infty, \quad a.s.\]
\end{defn}
%

\begin{defn}
A process on a probability space $(\Omega, {\cal F}, P)$ is asymptotically mean stationary (AMS) if there exists a probability measure $\bar{P}$ such that
\[\lim_{N \to \infty} {1 \over N} \sum_{k=0}^{N-1} P(T^{-k} F) = \bar{P} (F),\]
for all events $F$. Here $\bar{P}$ is called the stationary mean of $P$, and is a stationary measure.
\end{defn}
${\bar P}$ is stationary since, by definition $\bar{P}(F) = \bar{P}(T^{-1}F)$, for all events $F$ in the tail sigma field for the shift.
A cyclo-stationary process is AMS. See for example \cite{Gaarder}, \cite{GrayKieffer} or \cite{GrayProbabilit} (Theorem 7.3.1), that is $N-$stationarity implies the AMS property. 
Asymptotic mean stationarity is a very important property:
\begin{enumerate}
\item The Shannon-McMillan-Breiman Theorem (The Entropy Ergodic Theorem) applies to finite alphabet AMS sources \cite{GrayKieffer} (see an extension for a more general class \cite{Barron}). In this case, the ergodic decomposition of the AMS process leads to almost sure convergence of the conditional entropies.

\item Birkhoff's ergodic theorem applies for bounded measurable functions $f$, if and only if the process is AMS \cite{GrayKieffer}.
\end{enumerate}

Let \[F=\{x: \lim_{N \to \infty} {1 \over N} \sum_{i=1}^N f(T^{i} x) \quad \mbox{exists}.\}\]


It follows that for an AMS process $m(F)=1$, with $m$ being the stationary mean of the process. Birkhoff's Almost-Sure Ergodic Theorem states the following: If a dynamical system is AMS with stationary mean $m$, then all bounded measurable functions $f$ have the ergodic property, and with probability 1,
\[\lim_{N \to \infty} {1 \over N} \sum_{i=0}^{N-1} f(T^{i} x) = E_{m_x} [f], \quad x \in F,\]
where $E_{m_x}$ denotes the expectation under measure $m_x$ and $m_x$ is the resulting ergodic measure with initial state $x$ in the ergodic decomposition of the asymptotic mean (\cite{GrayProbabilit2} Theorem 1.8.2): $m(A) = \int m_x(A) m(dx)$. Furthermore,
\[\lim_{N \to \infty} {1 \over N} E[\sum_{i=0}^{N-1} f(T^{i} x) ]= E_{m} [f], \quad x \in F,\]

In fact, the above applies for all integrable functions (integrable with respect to the asymptotic mean).

\begin{defn} A random process is second-moment stable if the following holds:
$$\lim_{N \to \infty} {1 \over N} E[\sum_{m=0}^{N-1} (X_m(x))^2] < \infty$$
\end{defn}

\begin{defn} A random process is second-moment stable almost surely if the following limit exists and is finite almost surely:
$$\lim_{N \to \infty} {1 \over N} \sum_{m=0}^{N-1} (X_m(x))^2 < \infty$$
\end{defn}

\subsection{Stochastic Stability of Markov Chains and Random-Time State-Dependent Drift Criteria}\label{RandomDriftTheoremSection}

In this section, we review the theory of stochastic stability of Markov chains. The reader is referred to Meyn and Tweedie \cite{MCSS} for a detailed discussion. The results on random-time stochastic drift follows from Y\"uksel and Meyn \cite{YukMeynTAC2010}, \cite{YukselAllerton09}.

We let ${\bf \phi} =\{ \phi_t, t\geq 0\}$ denote a Markov chain with state space $\state$.
The basic assumptions of \cite{MCSS} are adopted:  It is assumed that $\state$ is a complete separable metric space,  that is locally compact; its  Borel $\sigma$-field is denoted $\clB(\state)$.  The transition probability is denoted by $P$, so that for any $\phi\in\state$,  $A\in\bx$,  the probability of moving in one step from the state $\phi$ to the set $A$  is given by  $  \Prob(\phi_{t+1}\in A \mid \phi_t=\phi) = P(\phi,A)$.   The $n$-step transitions are obtained via composition in the usual way, $  \Prob(\phi_{t+n}\in A \mid \phi_t=\phi) = P^n(\phi,A)$, for any $n\ge1$.   The transition law acts on measurable functions $f\colon\state\to\Re$ and measures $\mu$ on $\bx$ via,
\[
Pf\, (\phi)\eqdef \int_{\state} P(\phi,dy) f(y),\quad \phi\in\state,
\]
\[\mu P\, (A) \eqdef \int_{\state} \mu(d\phi)P(\phi,A) ,\quad A\in\bx.
\]
A probability measure $\pi$ on $\bx$ is called invariant if $\pi P= \pi$.  That is,
\[
\int \pi(d\phi) P(\phi,A) = \pi(A),\qquad A\in\bx.
\]


For any initial probability measure $\nu$ on $\bx$ we can construct a stochastic process with transition law $P$, and satisfying $\phi_0\sim \nu$.  We let $\Prob_\nu$ denote the resulting probability measure on sample space, with the usual convention for $\nu=\delta_{\phi}$ when the initial state is $\phi\in\state$. When $\nu=\pi$ then the resulting process is stationary.


There is at most one stationary solution under the following irreducibility assumption.  For a set $A\in\bx$ we denote,
\begin{equation}
\tau_A\eqdef \min(t \ge 1 :  \phi_t \in A)
\label{e:tauA}
\end{equation}
\begin{defn}
Let  $\varphi$ denote  a sigma-finite measure on $\bx$.
\begin{romannum}
\item
The Markov chain is called \textit{$\varphi$-irreducible} if for any $\phi\in\state$,
and  any $B\in\bx$ satisfying $\varphi(B)>0$,     we have
\[
\Prob_{\phi}\{\tau_B<\infty\} >0 \, .
\]

\item
A $\varphi$-irreducible Markov chain is \textit{aperiodic}  if  for any $\phi\in\state$, and any $B\in\bx$ satisfying $\varphi(B)>0$,   there exists $n_0=n_0(\phi,B)$ such that
\[
 P^n(\phi,B)>0 \qquad \hbox{\it for all \ } n\ge n_0.
\]

\item
A $\varphi$-irreducible Markov chain is \textit{Harris recurrent}  if   $\Prob_{\phi}(\tau_B < \infty  ) =1 $ for any $\phi\in\state$, and any $B\in\bx$ satisfying $\varphi(B)>0$.  It is \textit{positive Harris recurrent} if in addition there is an invariant probability measure $\pi$.
\end{romannum}
\end{defn}

Tied to $\varphi$-irreducibility is the existence of  {\em small} or {\em petite} sets. A set $A \in\bx$ is small if there is an integer $n_0\ge 1$  and a positive measure $\mu$ satisfying $\mu(\state)>0$ and
$$
P^{n_0}(\phi,B)  \geq \mu(B), \quad \hbox{\it for all \ }  \phi \in A, \; \mbox{and} \, B \in \bx .
$$
A set $A \in\bx$ is petite if there is a probability measure  $\sampd $ on the non-negative integers $\mathbb{N}$, and a positive measure $\mu$ satisfying $\mu(\state)>0$ and
$$
\sum_{n=0}^{\infty} P^n(\phi,B) \sampd (n) \geq \mu(B), \quad \hbox{\it for all \ }  \phi \in A, \; \mbox{and} \, B
\in \bx .
$$

\begin{thm}\label{MeynAnnals} [\cite{MCSS} Thm. 4.1]
Suppose that $\bfmX$ is a $\varphi$-irreducible Markov chain,
and suppose that there is a  set $A\in\bx$ satisfying the following:
\begin{romannum}
\item $A$ is $\mu$-petite for some $\mu$.
\item $A$ is \textit{recurrent}:    $\Prob_{\phi}(\tau_A < \infty  ) =1 $ for any $x\in\state$.
\item $A$ is {\em finite mean recurrent}:  $\displaystyle \sup_{\phi \in A} \Expect_{\phi}[ \tau_A] < \infty$.
\end{romannum}
Then $\bfmX$ is positive Harris recurrent.
\qed \end{thm}
Let $\stp_{z}, z \geq 0$ be a sequence of stopping times, measurable on a filtration generated by the state process with $\stp_{0}=0$.

\begin{thm}
\label{thm5} \cite{YukMeynTAC2010} \cite{YukselAllerton09}
Suppose that ${\bf \phi}$ is a $\varphi$-irreducible and aperiodic Markov chain.   Suppose moreover that there are functions
 $V \colon \state \to [0,\infty)$,
 $\delta \colon \state \to [1,\infty)$,
 $f\colon \state \to [1,\infty)$,
a small set $C$,   and a constant $b \in \Re$,   such that the following hold:
\begin{equation}
\begin{aligned}
\Expect[V(\phi_{\stp_{z+1}}) \mid \clF_{\stp_z }]  &\leq  V(\phi_{\stp_{z}}) -\delta(\phi_{\stp_z }) + b1_{\{\phi_{\stp_z} \in C\}}
\\
 \Expect \Bigl[\sum_{k=\stp_z}^{\stp_{z+1}-1} f(\phi_k)  \mid \clF_{\stp_z }\Bigr]  &\le \delta(\phi_{\stp_z})\, , \qquad \qquad \qquad \qquad z\ge 0.
\end{aligned}
\label{e:thm5delta}
\end{equation}
Then the following hold:
\begin{romannum}
\item
${\bf \phi}$ is positive Harris recurrent, with unique invariant distribution $\pi$
\item $\pi(f)\eqdef \int f(\phi)\, \pi(d\phi) <\infty$
\item  For any function $g$ that is bounded by $f$, in the sense that $\sup_{\phi} |g(\phi)|/f(\phi)<\infty$, we have convergence of moments in the mean, and the Law of Large Numbers holds:
\[
\begin{aligned}
\lim_{t\to\infty} \Expect_{\phi}[g(\phi_t)] &= \pi(g)
\\
\lim_{N\to\infty} \frac{1}{N} \sum_{t=0}^{N-1} g(\phi_t) &= \pi(g)\qquad a.s.\,, \ \phi\in\state
\end{aligned}
\]
\end{romannum}
\qed
\end{thm}

\begin{remark}
We note that the condition $f\colon \state \to [1,\infty)$ can be relaxed to $f\colon \state \to [0,\infty)$ provided that one can show that there exists an invariant probability measure.
\end{remark}

We conclude by stating a simple corollary to Theorem~\ref{thm5}, obtained by taking $f(\phi)=1$ for all $\phi \in \state$.
\begin{cor}\label{corol} \cite{YukMeynTAC2010} \cite{YukselAllerton09}
Suppose that ${\bf \phi}$ is a $\varphi$-irreducible Markov chain.   Suppose moreover that there is a function $V: \state \to (0,\infty)$, a small set $C$, and a constant $b \in \Re$,   such that the following hold:
\begin{equation}\label{PositiveRecur}
\begin{aligned}
\Expect[V(\phi_{\stp_{z+1}}) &\mid \clF_{\stp_z }] \leq V(\phi_{\stp_z }) - 1 + b1_{\{\phi_{\stp_z} \in C\}}
\\
\sup_{\ z\ge 0} \Expect[\stp_{z+1} -\stp_z &\mid \clF_{\stp_z }] < \infty.
\end{aligned}
\end{equation}
Then ${\bf \phi}$ is positive Harris recurrent.
\qed
\end{cor}

The following is a useful result for the paper. 
\begin{thm}\label{thm26} \cite{MCSS}
Without an irreducibility assumption, if \eqref{PositiveRecur} holds for a measurable set $C$, a function
 $V \colon \state \to (0,\infty)$, with $\sup_{x \in C} V(x) < \infty$, then $C$ satisfies $\sup_{x \in C} E[\tau_C]< \infty$.
\end{thm}

We have the following results. A Positive Harris Recurrent Markov process (thus with a unique invariant distribution on the state space) is also ergodic in the sense of ergodic theory (the ergodic theorem for Markov chains has typically a more specialized meaning with the state process being a coordinate process in the infinite dimensional space $\mathbb{X}^{\infty}$, see \cite{Hernandez}), which however, implies the definition in the more general sense. This follows from the fact that, it suffices to test ergodicity on the sets which generate the sigma algebra (that is the finite dimensional sets), which in turn can be verified by the recurrence of the individual sets; probabilistic relations in arbitrary finite sets characterize the properties in the infinite collection, and that, mixing leads to ergodicity.

%

\section{Acknowledgements}
 Discussions with Professors Sean P. Meyn, Robert M. Gray, Nuno Martins and Tam\'as Linder on the contents of the paper are gratefully acknowledged. The incisive reviews of two anonymous reviewers and the suggestions of the associate editor have led to significant improvement in the presentation.

\begin{IEEEbiographynophoto}
{Serdar Y\"uksel} received his BSc degree in Electrical and Electronics
Engineering from Bilkent University in 2001; MS and PhD degrees in
Electrical and Computer Engineering from the University of Illinois at
Urbana-Champaign in 2003 and 2006, respectively. He was a post-doctoral
researcher at Yale University for a year before joining Queen's University
as an assistant professor of Mathematics and Engineering at the Department
of Mathematics and Statistics. His research interests are on stochastic
and decentralized control, information theory and applied probability.
\end{IEEEbiographynophoto}
\end{document}

%% file: symbols.tex
\def\mindex#1{\index{#1}}

\def\sq{\hbox{\rlap{$\sqcap$}$\sqcup$}}
\def\qed{\ifmmode\sq\else{\unskip\nobreak\hfil
\penalty50\hskip1em\null\nobreak\hfil\sq
\parfillskip=0pt\finalhyphendemerits=0\endgraf}\fi\medskip}

\long\def\defbox#1{\framebox[.9\hsize][c]{\parbox{.85\hsize}{%
\parindent=0pt
\baselineskip=12pt plus .1pt      
\parskip=6pt plus 1.5pt minus 1pt 
 #1}}}

\long\def\beginbox#1\endbox{\subsection*{}%
\hbox{\hspace{.05\hsize}\defbox{\medskip#1\bigskip}}%
\subsection*{}}

\def\endbox{}

\newsavebox{\junk}
\savebox{\junk}[1.6mm]{\hbox{$|\!|\!|$}}

\def\limsup{\mathop{\rm lim\ sup}}
\def\liminf{\mathop{\rm lim\ inf}}

\def\state{{\mathbb X}}

\def\bx{{{\cal B}(\state)}}

\newcommand{\field}[1]{\mathbb{#1}}

\def\Re{\field{R}}

\def\bfmath#1{{\mathchoice{\mbox{\boldmath$#1$}}%
{\mbox{\boldmath$#1$}}%
{\mbox{\boldmath$\scriptstyle#1$}}%
{\mbox{\boldmath$\scriptscriptstyle#1$}}}}

\def\bfmX{\bfmath{X}}

\def\bfmY{\bfmath{Y}}

\def\bfmhhaY{\bfmath{\hhaY}} 
\def\bfmhhaY{\hbox to 0pt{$\widehat{\bfmY}$\hss}\widehat{\phantom{\raise 1.25pt\hbox{$\bfmY$}}}}

\def\til={{\widetilde =}}

\def\clB{{\cal B}}

\def\clF{{\cal F}}

 \def\FRAC#1#2#3{\genfrac{}{}{}{#1}{#2}{#3}}

\def\ddtp{{\mathchoice{\FRAC{1}{d^{\hbox to 2pt{\rm\tiny +\hss}}}{dt}}%
{\FRAC{1}{d^{\hbox to 2pt{\rm\tiny +\hss}}}{dt}}%
{\FRAC{3}{d^{\hbox to 2pt{\rm\tiny +\hss}}}{dt}}%
{\FRAC{3}{d^{\hbox to 2pt{\rm\tiny +\hss}}}{dt}}}}

\def\half{{\mathchoice{\FRAC{1}{1}{2}}%
{\FRAC{1}{1}{2}}%
{\FRAC{3}{1}{2}}%
{\FRAC{3}{1}{2}}}}

\def\eqdef{\mathbin{:=}}

\def\Prob{P}

\def\Expect{E}

\def\average#1,#2,{{1\over #2} \sum_{#1}^{#2}}

\def\eye(#1){{\bf(#1)}\quad}

\def\varble{\,\cdot\,}

\newtheorem{theorem}{Theorem}[section]

\def\Section#1{Section~\ref{#1}}

\def\eq#1/{(\ref{e:#1})}

\newcommand{\beqn}[1]{\notes{#1}%
\begin{eqnarray} \elabel{#1}}

\newcommand{\eeqn}{\end{eqnarray} }

\newcommand{\beq}[1]{\notes{#1}%
\begin{equation}\elabel{#1}}

\newcommand{\eeq}{\end{equation}}

\def\bdes{\begin{description}}
\def\edes{\end{description}}

\newcounter{rmnum}
\newenvironment{romannum}{\begin{list}{{\upshape (\roman{rmnum})}}{\usecounter{rmnum}
\setlength{\leftmargin}{14pt}
\setlength{\rightmargin}{8pt}
\setlength{\itemindent}{-1pt}
}}{\end{list}}

\newcounter{anum}

{\end{list}}

\def\ass(#1:#2){(#1\ref{#1:#2})}

\def\ritem#1{
\item[{\sf \ass(\current_model:#1)}]
}

\newenvironment{recall-ass}[1]{%
\begin{description}
\def\current_model{#1}}{
\end{description}
}

\def\Ebox#1#2{%
\begin{center}
 \parbox{#1\hsize}{\epsfxsize=\hsize \epsfbox{#2}}
\end{center}}

\newcommand{\bd}{\begin{description}}
\newcommand{\ed}{\end{description}}
\newcommand{\bt}{\begin{theorem}}
\newcommand{\et}{\end{theorem}}
\newcommand{\ba}{\begin{array}{rcl}}
\newcommand{\ea}{\end{array}}